\documentclass[american,aps,superscriptaddress,twocolumn,pra,aps,showpacs,floatfix,notitlepage]{revtex4-1}
\usepackage[T1]{fontenc}
\usepackage[latin9]{inputenc}
\usepackage{color}
\usepackage{babel}
\usepackage{amsthm}
\usepackage{amsmath, amssymb}
\usepackage{bm}
\usepackage{pxfonts,txfonts}
\usepackage{tgpagella}
\usepackage[T1]{fontenc}

\usepackage{amsmath}
\usepackage{amssymb}
\usepackage{graphicx}
\usepackage[unicode=true,pdfusetitle,
 bookmarks=true,bookmarksnumbered=false,bookmarksopen=false,
 breaklinks=false,pdfborder={0 0 0},backref=false,colorlinks=true]
 {hyperref}
\hypersetup{
 linkcolor=urlcolor,citecolor=urlcolor,urlcolor=urlcolor}

\makeatletter
\@ifundefined{textcolor}{}
{%
 \definecolor{BLACK}{gray}{0}
 \definecolor{WHITE}{gray}{1}
 \definecolor{RED}{rgb}{1,0,0}
 \definecolor{GREEN}{rgb}{0,1,0}
 \definecolor{BLUE}{rgb}{0,0,1}
 \definecolor{CYAN}{cmyk}{1,0,0,0}
 \definecolor{MAGENTA}{cmyk}{0,1,0,0}
 \definecolor{YELLOW}{cmyk}{0,0,1,0}
 }
\theoremstyle{plain}

\theoremstyle{plain}

\ifx\proof\undefined
\newenvironment{proof}[1][\protect\proofname]{\par
\normalfont\topsep6\p@\@plus6\p@\relax
\trivlist
\itemindent\parindent
\item[\hskip\labelsep
\scshape
#1]\ignorespaces
}{%
\endtrivlist\@endpefalse
}
\providecommand{\proofname}{Proof}
\fi
\theoremstyle{plain}

\theoremstyle{plain}

\usepackage{txfonts}
\definecolor{urlcolor}{rgb}{0,0,0.7}

\def \beq{\begin{equation}}
\def \eeq{\end{equation}}
\def \bse{\begin{subequations}}
\def \ese{\end{subequations}}
\def \bea{\begin{eqnarray}}
\def \eea{\end{eqnarray}}
\def \bem{\begin{displaymath}}
\def \eem{\end{displaymath}}
\def \bem{\begin{pmatrix}}
\def \eem{\end{pmatrix}}

\def \bs{\boldsymbol}
\def \nn{\nonumber}
\def \hh{\hat{H}}

\def \hb{\hat{b}}

\def \hc{\text{H.c.}}

\newcommand{\bra}[1]{\langle#1|}
\newcommand{\ket}[1]{|#1\rangle}

\makeatother

\providecommand{\lemmaname}{Lemma}
\providecommand{\propositionname}{Proposition}
\providecommand{\theoremname}{Theorem}
\providecommand{\corrolaryname}{Corrolary}

\begin{document}

\title{Modified spin-wave theory and spin liquid behavior of cold bosons on an inhomogeneous triangular lattice}

\author{Alessio Celi}

\affiliation{ICFO -- Institut de Ci\`encies Fot\`oniques, The Barcelona Institute
of Science and Technology, 08860 Castelldefels, Spain}

\author{Tobias Grass}

\affiliation{ICFO -- Institut de Ci\`encies Fot\`oniques, The Barcelona Institute
of Science and Technology, 08860 Castelldefels, Spain}

\author{Andrew J. Ferris}

\affiliation{ICFO -- Institut de Ci\`encies Fot\`oniques, The Barcelona Institute
of Science and Technology, 08860 Castelldefels, Spain}

\author{Bikash Padhi}

\affiliation{Department of Physics and Institute for Condensed Matter Theory, University of Illinois at Urbana-Champaign, Urbana, 61801 IL, U.S.A.}

\author{David Ravent\'os}

\affiliation{ICFO -- Institut de Ci\`encies Fot\`oniques, The Barcelona Institute
of Science and Technology, 08860 Castelldefels, Spain}

\author{Juliette Simonet}

\affiliation{Institut f\"ur Laserphysik, Universit\"at Hamburg, Luruper Chaussee 149, D-22761 Hamburg, Germany
}

\affiliation{Zentrum f\"ur Optische Quantentechnologien, Universit\"at Hamburg, Luruper Chaussee 149, D-22761 Hamburg, Germany
}

\author{Klaus Sengstock}

\affiliation{Institut f\"ur Laserphysik, Universit\"at Hamburg, Luruper Chaussee 149, D-22761 Hamburg, Germany
}

\affiliation{Zentrum f\"ur Optische Quantentechnologien, Universit\"at Hamburg, Luruper Chaussee 149, D-22761 Hamburg, Germany
}

\author{Maciej Lewenstein}

\affiliation{ICFO -- Institut de Ci\`encies Fot\`oniques, The Barcelona Institute
of Science and Technology, 08860 Castelldefels, Spain}

\affiliation{ICREA -- Instituci\'o Catalana de Recerca i Estudis Avan\c cats, Lluis
Companys 23, 08010 Barcelona, Spain}

\email{alessio.celi@icfo.es}

\begin{abstract}
Ultracold bosons in a triangular lattice are a promising candidate for observing quantum spin liquid behavior.
Here we investigate, for such system, the role of a harmonic trap giving rise to an inhomogeneous density. 
We construct a modified spin-wave theory 
for arbitrary filling, and predict the breakdown of order for certain values of the lattice anisotropy. 
These regimes, identified with the spin liquid phases, are found to be quite robust upon changes in the filling factor. 
This result is backed by an exact diagonalization study on a small lattice.
\end{abstract}


\maketitle

\section{Introduction}

Quantum spin liquids (QSL) are at the center of interest of
contemporary condensed matter physics and quantum many body theory
(cf. \cite{Balents2010}) for several reasons. P. W. Anderson
proposed them as a new kind of insulator: a resonating valence
bond (RVB) state \cite{Anderson1973}. The interest in these state
was clearly stimulated by the fact that they were soon associated
with high $T_c$ superconductivity \cite{Anderson1987}. Immediately
it was realized that RVB spin liquids  might exhibit topological
order \cite{Kivelson1987} and are related to fractional quantum
Hall states \cite{Kalmeyer1987} and chiral spin states
\cite{Wen1989}.

Frustrated anti-ferromagnets (AFM) provide  paradigm playground
for RVB states and spin liquids (for the early reviews see
\cite{Misguich2003,Lhuillier2005,Alet2005}). The most prominent
example is Heisenberg spin $1/2$ model in a kagome lattice.
Unfortunately, they are notoriously difficult for numerical
simulations, since due to the (in)famous sign problem quantum
Monte Carlo methods cannot be applied. Still, a lot of information
can be extracted from exact diagonalization studies (for seminal
early studies see Ref. \cite{Waldtmann1998}). There was a lot of
effort to describe QSLs with various approximate analytic
approaches, such as large $N$ expansion \cite{Read1991},  or
appropriate mean field theory \cite{Wen1991,Wen2004}. These
studies suggested that QSLs described by RVB states represent
topologically ordered states with finite energy gap, analogous to
those of the famous Kitaev's Toric Code model \cite{Kitaev2003}.

In parallel to AFM in kagome lattice, the so called  dimer model
in triangular lattice was studied intensively \cite{Moessner2001}
-- it was also found that it is expected to exhibit a gapped RVB
phase (see also \cite{Moessner2002,Montrunich2005}).

The first experimental indications of QSLs comes from studies of
Mott insulator in the triangular lattices \cite{Shimizu2003}, and
power law conductivity inside the Mott gap in certain materials
\cite{Ng2007}. More recently observations (cf.
\cite{Pratt2011,Han2012,Fu2015}) combine various standard and
non-conventional detection methods in kagome Heisenberg AFM,
including measurements of fractionalized excitations
\cite{Han2012}. There are also reports of QSL behavior in the, so
called,  {\it Herbertsmithites} (cf. \cite{Amusia2014}).

Recently great progress was achieved in numerical simulations of
the gapped QSLs, based on the the use 1D density matrix
renormalization group (DMRG) codes, ``wired'' on 2D tori/cylinders.
This approach allowed for better insight into the properties of
the ground state of the Heisenberg AFM in the kagome lattice
\cite{Yan2011,Depenbrock2012}. More importantly, it allowed
obtaining  convincing signature of its topological $Z_2$ nature
\index{Jiang2012}. This was  based on numerical estimate for the,
so called, topological entanglement entropy (TEE) -- the quantity
that unambiguously  characterizes topological gapped QSLs
\cite{Levin2006,Kitaev2006}. Calculations of TEE were earlier
applied to the quantum dimer model in the triangular lattice
\cite{Furukawa2007} and to the Bose-Hubbard spin liquid in the
kagome lattice \cite{Isakov2011}. They were extended to critical
QSLs \cite{Zhang2011}, Toric Code \cite{Jiang2012} and lattice
Laughlin states \cite{Zhang2011a}. Since these calculations aim at
sub-leading term in entanglement entropy,  it is  quite
challenging to achieve good accuracy (see for instance
\cite{Jiang2012,Zhang2013}.

Recently, studies of AFM in kagome lattice were extended to novel
proposals for characterizing/detecting topological excitations
and dynamical structure factor \cite{Punk2014}. Several papers
discuss inclusion of the chiral terms and Dzyaloshinsky-Moriya
interactions, resulting in formation of chiral QSLs
\cite{Wietek2014,Kumar2015}. Considerable interest was devoted
also to the $J_1-J_2$ Heisenberg model in the kagome lattice
\cite{Kolley2015} and in the square lattice \cite{Jiang2012a}, 
to the $J_1-J_2-J_3$ Heisenberg model in the kagome lattice \cite{Gong14,Gong15}
, and to the Kitaev-Heisenberg model \cite{Kitaev06,Kimchi14} in triangular lattices
\cite{KaiLi15,Rousochatzakis16}%
.

Systems of ultracold atoms and ions provide a very versatile
playground for quantum simulation of various models of theoretical
many body physics \cite{Lewenstein2007,Lewenstein2012} -- QSL have
in this context also quite long history. The first proposals for
quantum simulators of the Kitaev model in the hexagonal lattice
\cite{Duan2003}, and AFM in the kagome lattice
\cite{Santos2004,Damski2005,Damski2005a} were formulated more than
ten years ago; all of them were based on smart designs and use of
super-exchange interactions in optical lattices. More feasible and
perhaps are experimentally less demanding proposals based on
ultracold ions \cite{Schmied2008}, or ultracold atoms in shaken
optical lattice \cite{Eckardt2010}. The latter schemes were
originally designed to control the value and sign of the tunneling
in Bose-Hubbard models -- for original theory proposal see
\cite{Eckardt2008}, and for the first experiments in the square
lattice see \cite{Zenesini2009}. They should be regarded as
specific examples of  generation of synthetic gauge fields in
optical lattices \cite{Lewenstein2012,Goldman2014}, or more
precisely synthetic gauge fields in periodically-driven quantum
systems \cite{Goldman2014a}.

Change of sign of tunneling in the triangular lattice is know to
be equivalent of the introduction of the $\pi$-flux synthetic
``magnetic'' field in the Bose-Hubbard model
\cite{Kalmeyer1987,Eckardt2010}. In the hardcore boson limit one
obtains then an XX AFM model in the triangular lattice, which
for isotropic bonds is known to have  a planar N\'eel ground
state. If, however, the bonds are anisotropic and their values $t_1$, 
$t_2=t_3=t\, t_1$ can be
controlled, then as anisotropic parameter $t$ goes from infinity to zero the model
interpolates between an AFM in a rhombic lattice (with the
conventional N\'eel ground state) to an AFM in the ideal
triangular lattice (with the planar N\'eel ground state), and
finally to an AFM in an array of weakly coupled 1D chains (with the
conventional N\'eel ground state again). Exact diagonalizations
and tensor network states simulations (PEPS) indicate that between
these three N\'eel phases there exists two quite extended regions
of gapped QSL \cite{Schmied2008}.

Interestingly the presence and the location of the QSL phases can
be determined quite accurately using the generalized spin wave
theory, which signals instability at the QSL boundaries
\cite{Schmied2008,Hauke2010}. The spin wave method is impressively
powerful and has been generalized and applied to frustrated bosons
and Heisenberg model with completely asymmetric triangular lattice
\cite{Hauke2011,Hauke2013a}.

We should stress that the proposal of Ref. \cite{Eckardt2010} is
in principle very promising, since it requires temperature of
order of $(t/U) U \simeq t$ which is achievable in realistic experimental
conditions, here $U$ denotes atom-atom on site interaction energy.
In fact, initial experiment demonstrated feasibility of the
scheme, but were conducted far from hardcore boson limit. 
 In these experiments a triangular  array of cigar shaped
Bose-Einstein condensates was realized, corresponding to a
frustrated quasi-classical AFM \cite{Struck2011}, described by a
classical XX spin model with the $U(1)$ symmetry, and Gaussian
Bogoliubov-de Gennes quantum, or better to say quasi-classical
fluctuations. In the further works, by exploiting control over the
temporal shape of the periodic modulation, one could realize
arbitrary Peierl's phases, i.e. arbitrary fluxes of the synthetic 
 ``magnetic'' field through the elementary plaquette of the lattice
(\cite{Struck2012}, see also \cite{Hauke2012}). This allowed for
realization of a quasi-classical spin model with competing $U(1)$
and Ising $Z_2$ symmetries \cite{Struck2013}. The route toward the
strongly correlated regime and hardcore limit seem to be obscured,
however, by uncontrolled heating mechanisms, most probably
intrinsically associated with the periodic modulation scheme
\cite{Goldman2014a}.

Even if this difficulty is overcome, another experimental aspect 
might prevent the observation of QSL in such systems. 
Indeed the overall harmonic trapping of the atomic ensemble leads
 to non-constant filling factor over the optical lattice. 
 We should expect thus formation of wedding cake structure, formed 
 by the different quantum phases (cf. \cite{Lewenstein2012}
and references therein). How does the phase diagram look
like or change in the presence of such ``experimental imperfections''?
This is the question we want to answer in this paper. To this aim
we apply exact diagonalization on small lattices , and wired DMRG
with open boundary conditions. On large lattices we apply modified
spin-wave theory, adopted to the spatially inhomogeneous
situation, which turns out
 to be technically much more  demanding than the one pertaining to the spatially homogeneous
 lattice. Our work provides a  starting point for the future
 applications of tensor network state approaches like Projected Entangled Pair States (PEPS) to a
 moderate size lattices. These future calculations will aim at
 estimations of topological entropy, which so far for the
 considered model in the triangular lattice has not yet been
 accomplished even in the spatially homogeneous case with periodic
 boundary conditions. Studying the influence of the spatial
 inhomogeneities, induced by the presence of the trap or
 disorder, on topological entanglement entropy is a fascinating
 question in itself -- it goes, however, beyond the scope of the
 present paper.
While inhomogeneity due to confinement are instrinsict to ultracold atoms,
our approach may be also relevant for searching QSLs in other quasi-2D condensed matter systems 
that present residual magnetization or inhomogeneities, for instance, due to the presence of a substrate.

The paper is structured in the following way: After introducing the system and model in Section II, 
we construct the modified spin wave theory in Section III. 
From this theory, we obtain a phase diagram in Section IV. 
In Section V, we consider first a small lattice using exact diagonalization. 
Then, we show that quasi-exact results can be obtained for much larger lattices using DMRG. 
The main conclusion drawn from our study, summarized in Section VI, 
regards the co-existence of spin liquid behavior at different filling factors smaller than 1/2. 
Thus, the spin liquid phase is expected to be robust against inhomogeneities due to a trapping potential. 
Our finding should facilitate the experimental observation of spin liquids in optical lattice systems.

\section{Description of the atomic model and map to the spin model}

Ultracold bosons in deep optical lattices are very well described by the Bose-Hubbard model. 
Therefore, we will take the Bose-Hubbard Hamiltonian as a starting point for our analysis:
\beq
\hh = \sum_{\langle i j \rangle} t_{ij} (\hb_i^\dag \hb_j + \hc) + \frac{U}{2} \sum_i \hat{n}_i (\hat{n}_i - 1) + \sum_i V_i \hat{n}_i .
\eeq
Here, the $\hb_i^\dag$, $\hb_i$ are the creation and annihilation operator at the site $i$ of the triangular lattice, and $\hat{n}_i= \hb_i^\dag \hb_i$ is the number 
operator of the Fock space.
The first term is a possibly anisotropic nearest-neighbor hopping, with tunneling amplitudes $t_{ij}$. 
In the standard case, one would have a minus sign in front of the tunneling term. 
However, it is possible to control the sign (or even phase) of the tunneling, 
which is a crucial ingredient to generate frustration in the triangular lattice. 
As here we will exclusively be interested in such scenario of reversed hopping amplitude, 
we absorbed the sign in the definition of $t_{ij}$, such that standard hopping would correspond to $t_{ij}<0$, while we will consider $t_{ij}>0$. 
The second term in $H$ describes repulsive on-site interactions of strength $U>0$.  
The last term is the trapping potential $V_i = V r_i^2 - \mu_0 , V = \frac{1}{2} m \omega^2$. 
Although it is present in any realistic experiment, it is often neglected in theoretical studies. 
The positions of a boson on site $i$ is denoted by $ r_i$. 

\begin{figure}
  \includegraphics[width=0.9\columnwidth, height=0.7\columnwidth]{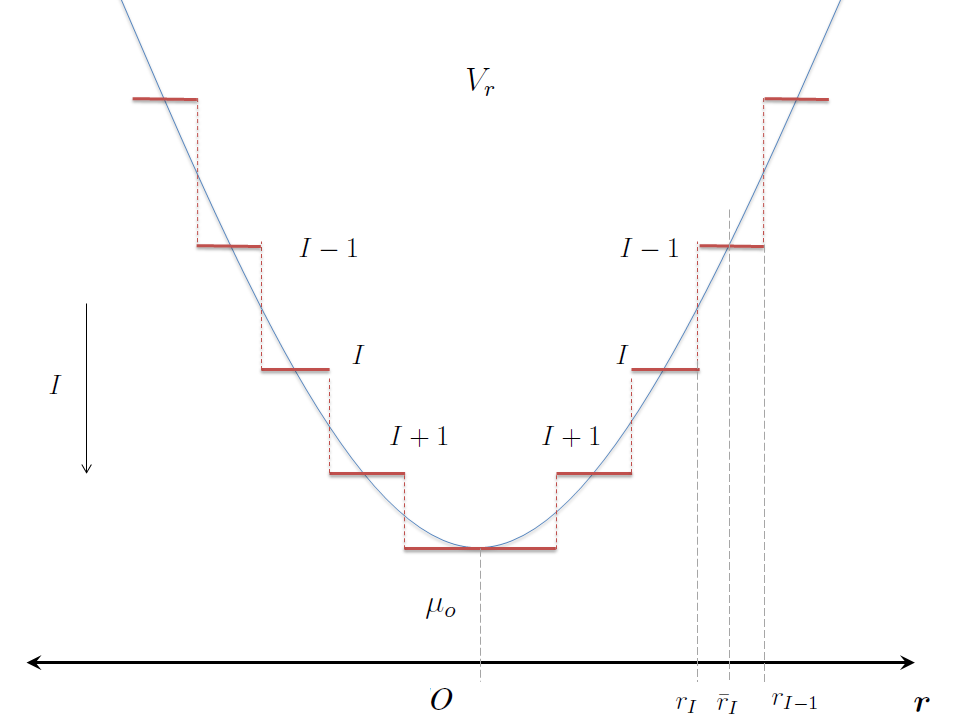}
 \caption{(Color online) 
Local density approximation. The harmonic potential, which for simplicity we assume to have cylindrical symmetry,  
is decomposed in two contributions: a step-like profile and a smoothly varying term.
Each plateau extends between the radii $r_I$ and $r_{I-1}$, defined as the distances 
where the average occupation takes two consecutive integer values, 
$\langle\hat n_{r_I}\rangle =I$, $\langle\hat n_{r_{I-1}}\rangle =I-1$. The hight of the plateau is taken to be the one corresponding to half filling. 
The smooth terms can be then treated as a perturbation, on the same footing as the hopping term. 
The effective model in each plateau  is thus equivalent to an anisotropic XX-spin model with
an smoothly varying magnetic term.}\label{fig:trap}
 \end{figure}

If interactions $U$ are strongly repulsive, fluctuations in particle number is  suppressed. 
It is then justified to restrict the local Hilbert space to a subspace formed by the states with occupation number two. 
These states may change throughout the trap, but within a local density approximation, 
we may keep them fixed within a circular area in the center of the trap, and ring-shaped areas further outside, as illustrated in Figure \ref{fig:trap}. 
Each region is denoted by an integer $I$, according to the possible occupation within the region, $n_I=\{I-1,I\}$.

This approach allows to map the Bose-Hubbard Hamiltonian onto a spin model, using a Holstein-Primakoff transformation \cite{Nolting}. 
Within each region $I$, the transformation is defined as
\bea
\hat{S}^z_i &=& (-1)^I \Big ( I - \frac{1}{2} - \hat{n}_i \Big ),  \nn \\
\hat{S}^+_i &=& \frac{(\hb^\dag)^I}{\sqrt{I}} \quad , \quad
\hat{S}^-_i = \frac{(\hb)^I}{\sqrt{I}} \quad , \quad (\hb^\dag)^2 = (\hb)^2 = 0.
\eea
The vanishing of squared creation or annihilation operators is due to the restriction of the local Hilbert space to two states.
Using the definition of spin operators the tunneling part of the original tight-binding Hamiltonian 
gets transformed to $I \sum_{\langle i,j \rangle} t_{ij} \hat{S}^+_i \hat{S}^-_{j} + \hc$. 
The interaction part transforms to $U(\hat{S}^z)^2 + 2U(-1)^{I +1} \hat{S}_z (I-1) + U(I^2 -2I + 3/4)$. The trap potential gives rise to a term $V_i \hat{S}^z_i$.
With $(\hat{S}^z)^2=1/4$, and neglecting the terms which are constant within a given region $I$, 
the dynamical part of the transformed Hamiltonian is an XX model in an inhomogeneous transverse field:
\beq
\hh_I = I \sum_{\langle i j \rangle} t_{ij} \hat{S}^+_i \hat{S}^-_j + \hc + \sum_i V_i \hat{S}_i^z .
\label{eq:H1}
\eeq

 \begin{figure}
  \includegraphics[width=0.92\columnwidth]{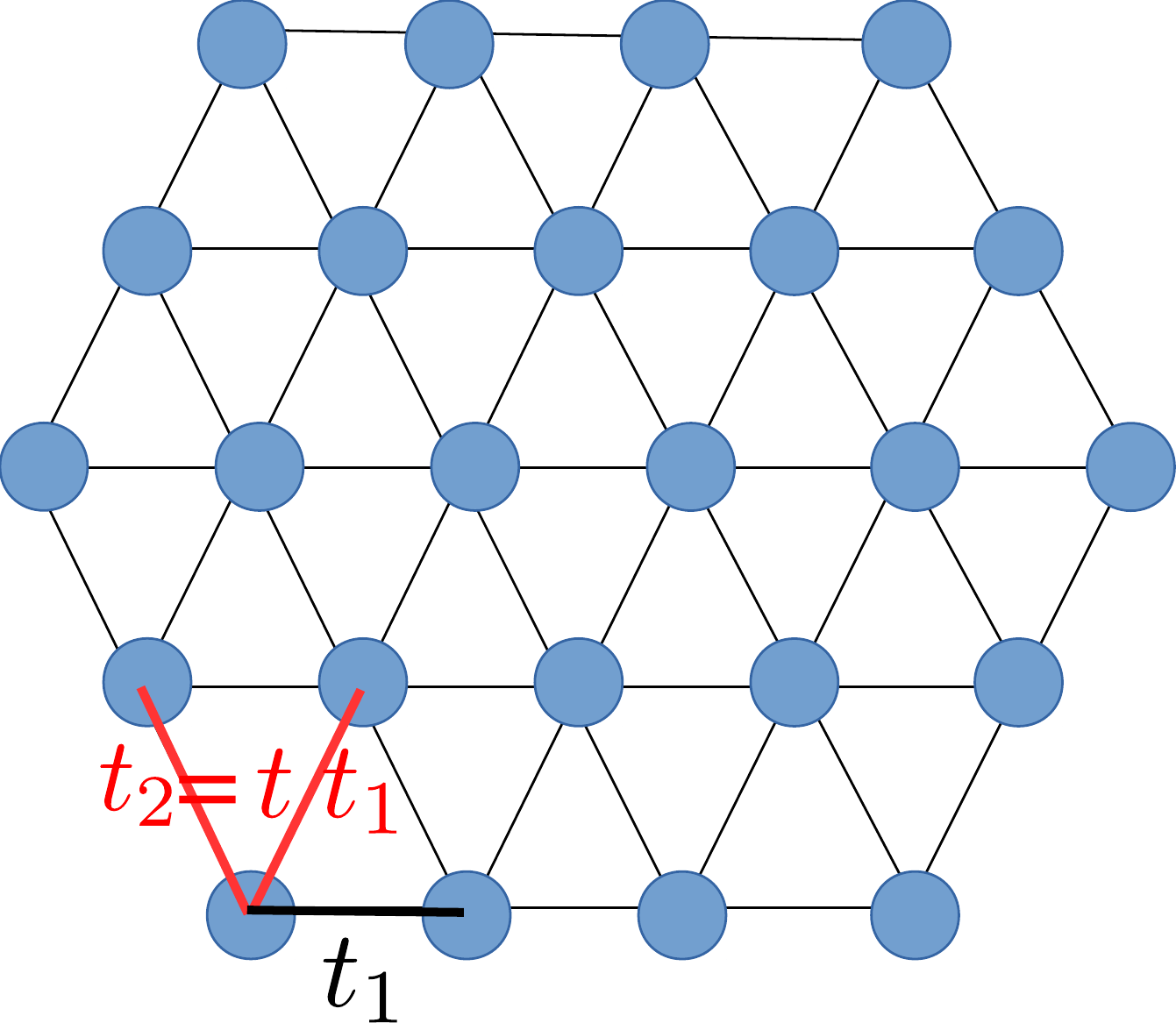}
\caption{(Color online) A triangular lattice of $N=24$ sites with a hexagonal shape. Horizontal hopping amplitudes 
are given by $t_1$, while hopping in the other directions have an amplitude $t_2= t\, t_1$, where $t$ parametrizes
 the anisotropy of the lattice.} \label{fig:lattice}
 \end{figure}
 
Before studying this Hamiltonian in the next sections using modified spin wave theory, exact diagonalization, and DMRG, 
let us briefly discuss the parameter regimes which are of interest experimentally. 
As mentioned before, being interested in frustration and spin liquids, the spin-spin interactions in Eq. (\ref{eq:H1}) 
should be antiferromagnetic, that is, $t_{ij}>0$. To simplify the scenario, $t_{ij}$ should only depend on the direction of the hopping, 
with amplitudes along horizontal links denoted $t_1$, while the two links with non-zero vertical component shall have an amplitude $t_2= t\,  t_1$, 
see Figure \ref{fig:lattice}. The anisotropy of the lattice is then characterized by a single parameter $t$, which we will tune from 0, 
corresponding to an effective 1D system, to values greater than $2$, where the lattice geometry is dominated by a rhombic structure. 
The energy difference between neighboring spins is of the order $\Delta V_i = V a^2 \equiv \eta\, t_1$, where $a$ is the lattice constant, 
and $\eta$ is a dimensionless parameter. We Assume the lattice is loaded
with $^{87}$Rb atoms, which has lattice constant $a = 553$ nm and a
trap frequency $\omega=2\pi \times 40$Hz; we have $\Delta V_i / \hbar= 15$Hz. 
This is about an order of magnitude weaker than typical interactions strengths, $t_1/\hbar \approx 150$Hz. 
In the modified spin wave approach, we will therefore take $V_i=0$, while the effect of non-zero values will be addressed within the exact diagonalization study.

\section{Modified spin-wave theory}

Let us start by analyzing the spin system for constant {\it non-zero} magnetization, which corresponds to fillings different from $\frac 12$. 
Classically, we expect the spin oriented along a cone around the $z$-axis,
\begin{equation}
{\bf S}_i=(\sin \rho\, \cos({\bf Q}\cdot {\bf r}_i),\sin \rho\, \cos({\bf Q}\cdot {\bf r}_i), \cos \rho). \label{eq:classord}
\end{equation}
Here, $\bs Q =(Q_x,Q_y)$ is a vector in the $xy$-plane while $\rho$ is the azimuthal angle related to the magnetization along the $z$-axis, i.e. to the filling
of the original bosons $\nu = \langle \hat n\rangle - [\langle \hat n\rangle]$, where $[x]= $ integer part of $x$.
 For $\rho= \frac {\pi}2$, \eqref{eq:classord} reduces to the ansatz considered by \cite{Xu1991,Hauke2010} 
at half filling. 
If we follow the standard spin-wave approach, we should choose the local basis in such a way that the new local $z$-axis is parallel to the vector \eqref{eq:classord}. 
In this way, by applying the bosonization of the local spin we would model fluctuations {\it along the classical ordering} represented by \eqref{eq:classord}.
Now, such fluctuations would have component also along the $z$-axis. In other words, they would renormalize the filling factor. 
Such behavior is not acceptable from the physical point of view. Indeed, in the original bosonic Hubbard model the filling factor is a well defined quantity: the hopping term conserves 
the particle number, and, thus, the expectation value of the particle density which is the filling. The same argument holds for the same physical model as described as a spin system.
In practice, the acceptable fluctuations are restricted to the $xy$-plane, and, precisely, are along the projection of the ordering vector on the $xy$-plane. That is to say that corrected 
choice for the quantization axis is the same as at half filling. 

What is then the difference with respect to the half-filling case? The difference resides in the magnitude of the spin projection. If we do the reasonable assumption that 
the fluctuations are proportional to such length we can relate $n$, the local density of bosonic excitations,  to the filling. As originally proposed by Takahashi \cite{Takahashi1989}, such density 
at half filling should be taken equal to the total spin, $n=S$, that to say also the bosonic excitations are at half-filling. Here, we propose a generalized Takahashi condition,
\begin{equation}
n=S |\sin \rho|,
\end{equation}
where the angle $\rho$ is related to the filling by the 
relation $ \langle S^z \rangle =\nu-S = S \cos \rho$ which implies $|\sin \rho|=\frac 1S \sqrt{\nu(2S-\nu)}$.
This choice has further physical justification. First, it is symmetric around half filling as it should be: reversing the quantization axis $\hat z$ 
in the Dyson-Maleev  transformation \cite{Dyson1956,Maleev1957,Maleev1958} is equivalent to the replacement $\nu \to 2S - \nu$. Second, fluctuations are maximal at half-filling 
and are suppressed in the paramagnetic (Mott) phases, which correspond to filling $\nu=0$ and $2 S$.

As derived in previous sections, the filling factor $\nu$ is smoothly changing in the trap and relates to the harmonic potential as $\nu= [\frac{\mu}U +\frac 12]$, 
where $[x]=$ fractional part of $x$ (the hopping term has zero mean). Thus, our analysis can be applied in local density approximation to trapped systems.

We define the local spin operators $\hat {\bf S}'\equiv (\hat S^{x'},\hat S^{y'},\hat S^{z'})$ 
that are related to the global ones $\hat {\bf S}= (\hat S^{x},\hat S^{y},\hat S^{z})$ through the rotation
\begin{equation}
\hat {\bf S} = R(\theta_i) \hat {\bf S}'  \equiv R({\bf Q}\cdot {\bf r}_i)  \hat {\bf S}',
\label{eq:rot}
\end{equation}
where 
$$R(\theta_i)= R_z(\theta_i) R_y(-\pi/2)R_z(\theta_i)
            =\left(\begin{matrix}
             0 & -\sin (\theta_i ) & -\cos (\theta_i ) \\
             0 & \cos (\theta_i ) & -\sin (\theta_i ) \\
              1 & 0 & 0 \\
                   \end{matrix}\right),$$ 
is the rotation that sends the vector $(0,0,-1)$ to $(\cos \theta_i,\sin \theta_i,0)$, i.e. along the projection of ordering vector on the $xy$-plane.

By composing with the Dyson-Maleev transformations
\begin{align}
\hat S_i^{z'} &\to -S + a_i^\dag a_i, \cr 
\hat S_i^{+'} &\to \sqrt{2S} a_i, \cr 
\hat S_i^{z'} &\to \sqrt{2S}(1 - \frac{a_i^\dag a_i}{2S}) a_i, \label{eq:DM}
\end{align}
we find 
that in the original spin basis, the bosonization is

\begin{align}
S_i^\pm &= e^{\pm i \theta_{i}} \left( \pm \sqrt{\frac S2} ( a_i^\dag - (1-\frac{\hat n_i}{2 S}) a_i) + S  (1-\frac{\hat n_i}{S})\right), 
\end{align}
where $\hat n_i = a^\dag_i a_i$,  $\theta_{i j} = \bs Q \cdot \bs r_{ij}$, and $\bs r_{ij} = \bs r_j - \bs r_i$. 
The effective Hamiltonian reads (up to fourth order in $a$ or $a^\dagger$)
\begin{widetext}
\begin{align}
H&= \frac 12 \sum_{<ij>} t_{ij} \left( S_i^+ S_j^- + S_i^- S_j^+ \right) 
\cr
  &= \sum_{<ij>} t_{ij} \cos \theta_{ij} \left( S^2   - S  (\hat n_i + \hat n_j) - \frac S2 (a_i^\dag a_j^\dag + a_i a_j ) + \frac S2  (a_i^\dag a_j + a_i a_j^\dag) \right. \cr
   & \left. + \hat n_i \hat n_j - \frac 14 (a_i^\dag\hat n_j    a_j  + a_j^\dag\hat n_i a_i )  + \frac 14  (\hat n_j a_j a_i  + \hat n_i a_i a_j)
 \right) \cr
  & - i \sum_{<ij>} t_{ij} \sin \theta_{ij} \left(   S \frac{\sqrt{2S}}2  (a_i^\dag - a_j^\dag - a_i +a_j ) +\frac{\sqrt{2S}}{4} (\hat n_i a_i -\hat n_j a_j ) 
  -\frac{\sqrt{2 S}}{2}  (\hat n_j  a_i^\dag - \hat n_i a_j^\dag -  \hat n_j a_i + \hat n_i a_j )   \right).   \label{heff}
\end{align}
\end{widetext}

Note that this expression does not coincide with \cite{Hauke2010}[Eq. 5]: indeed, the odd terms in $\sin \theta_{ij}$ are absent there as they have zero expectation value on a thermal gas of excitations.
It is worth noticing that the terms in $\cos\theta_{ij}$ and $\sin \theta_{ij}$  are manifestly symmetric and antisymmetric under the exchange of indices, $i\leftrightarrow j$, respectively. 
Indeed, by construction the whole expression is invariant under such exchange  of summed indices. Furthermore,
the Hamiltonian \eqref{heff} can be rewritten in an explicit translational invariant fashion by noticing that the sum over the links can be performed as a sum over there links coming out of a site, 
and then summing over all the sites. As these three lattice vectors on a triangular lattice we choose $ \bs \tau_1=(1,0), \bs \tau_2 = \frac 12 (1, \sqrt 3), \bs \tau_3 = \frac 12(-1, \sqrt 3)$.
 As $H$ in Eq. \eqref{heff} is non-Hermitian, following Takahashi \cite{Takahashi1989}, 
we use it in order to construct a free Energy for a gas of bosonic  excitations in a generic Bogoliubov basis  at temperature $T$, i.e.
\begin{equation}
{\cal F} = E-T \mathcal{S} + \mu (n-S |\sin \rho|),\label{freeenergy}
\end{equation}
where $E$ is the expectation value of $H$,
\begin{equation}
E=\frac 1N \sum_{\bs k} \bra{\nu_{\bs k}} H\ket{\nu_{\bs k}}, \ \ \nu_{\bs k}\equiv \langle \alpha_{\bs k}^\dag \alpha_{\bs k}\rangle =\frac 1{\exp[w_{\bs k}/T] -1},
\end{equation}
with $\alpha_{\bs k}$ denoting the Bogoliubov modes, see Eq. ( \ref{bgtra0} ). The entropy $\mathcal{S}$ of the bosonic gas is defined as
\begin{equation}
\mathcal{S} = \frac 1N \sum_{\bs k} [ (\nu_{\bs k} + 1) \ln (\nu_{\bs k} +1) - \nu_{\bs k} \ln \nu_{\bs k} ].
\end{equation}
The last term in Eq. (\ref{freeenergy}) is the modified Takahashi constraint over the density of fluctuations $n= \langle \hat n_i \rangle$, with $\mu$ the corresponding Lagrange multiplier or chemical potential.
Here, $w_{\bs k}$ is energy of each mode. From the functional form of the entropy it follows that $w_{\bs k}$ is also the rate at which the entropy changes with changing occupation, i.e.
$w_{\bs k} = T \frac{\partial \mathcal{S}}{\partial \nu_{\bs k}}$.

It seems natural to adopt this strategy since the expectation value $E$ is in general bounded from below and depends only on the average value of the bilinears $a_i^\dag a_j$, $a_i^\dag a_j^\dag$, and their complex conjugates. 
This happens because the Bogoliubov transformation is by definition linear and only the quadratic bilinears above can have non-zero matrix elements while preserving the excitation number. 
This physical consideration is equivalent to state that $E$ can be calculated using Wick theorem and that linear and cubic terms give zero contribution. For convenience, we define
\begin{align}
\langle a_i^\dag a_j\rangle &\equiv F(r_{ij}) -\frac 12 \delta_{ij},\cr
\langle a_i a_j\rangle &\equiv G(r_{ij}) .
\end{align}
In  this notation, the generalized Takahashi constraint reads
\begin{equation}
F(0)= \langle a^\dag a \rangle + \frac 12 = S |\sin \rho| + \frac 12, \label{eq:gentaka}
\end{equation}  
where $|\sin \rho|$ relates to the filling $\nu$ of the original spin system, $0\le\nu\le 2S$, as $|\sin \rho|=\frac 1S \sqrt{\nu(2S-\nu)}$. 
From \eqref{heff} we find
\begin{align}
\frac EN &= S^2 \;C  - 2 S\;  C  \left[ F(0) -\frac 12 \right] - \frac S2 \sum_J \left[ c_J \cdot ( G_J +  G^*_J -  F_J  -  F^*_J ) \right]  \cr
         &\,+ C  \left[ F(0)-\frac 12 \right]^2 +\sum_J c_J (|F_J|^2 + |G_J|^2)\cr
         &\, + \frac 14  c_J \left\{ (G(0)  (F_J + F_J^* - 2 G_J^*) - 2 \left[ F(0)- \frac 12 \right]  (F_J + F_J^* - 2 G_J) \right\} .\label{E}
         \end{align}
Here, we adopt the notation $F_J \equiv F(\tau_J)$, $G_J \equiv G(\tau_J)$, and we define $(c_1,c_2,c_3)\equiv (\cos (\bs Q \cdot \tau_1),t \cos (\bs Q \cdot \tau_2),t \cos (\bs Q \cdot \tau_3))$, 
$C \equiv c_1 +c_2 +c_3$. For convenience, we fix the energy scale such to have $t_1$. 

If we assume that the Bogoliubov transformation is real as in \cite{Hauke2010} 
\begin{align}
 a_{\bs k} &= \big ( \cosh \theta_{\bs k} \;\alpha_{\bs k} + \sinh \theta_{\bs k} \; \alpha^\dag_{-\bs k} \big )   \, , \cr
a_{-\bs k} &= \big ( \cosh \theta_{\bs k} \; \alpha_{- \bs k} + \sinh \theta_{\bs k} \; \alpha^\dag_{\bs k} \big )   \, , \label{bgtra0}
\end{align}
we have that  $F_J= F_J^*$, $ G_J= G_J^*$, the expectation value of energy density reduces to
\begin{align}
\frac EN &= \frac 12 \sum_J c_J \left[\left(S  + \frac 12  - F(0) + F_J \right)^2 + \left(S  + \frac 12  - F(0)- G_J \right)^2 \right.\cr 
&\ \left. + G(0)(F_J-G_J) + F_J^2+G_J^2\right], \label{eq:Ebg}
\end{align}
which differs from the expression \cite[Eq.6]{Hauke2010} not only due to the mismatch between our \eqref{heff} and \cite[Eq.5]{Hauke2010}: in fact the term $G(0)(F_J-G_J)$ is omitted as considered negligible.
This approximation is justified at half filling for large $S$ as $F_J \sim G_J \sim S$.  

It is worth noticing that the structure of the minimal solution is not affected by the explicit form of $E$, while the consistency equations obviously are.
Indeed, due to \eqref{bgtra0} the expectation values have the form
\begin{align}
F(\bs r) &=  \frac 1N \sum_{\bs k} \cosh (2\theta_{\bs k}) e^{-i \bs k \bs r} \left(\nu_{\bs k} + \frac 12 \right)\cr
         &=\frac 1N \sum_{\bs k'} \cosh (2\theta_{\bs k'}) \cos(\bs k' \bs r) (2\nu_{\bs k'} + 1), \cr
G(\bs r) &=  \frac 1N \sum_{\bs k} \sinh (2\theta_{\bs k}) e^{-i \bs k \bs r} \left(\nu_{\bs k} + \frac 12 \right)\cr
         &=\frac 1N \sum_{\bs k'} \sinh (2\theta_{\bs k'}) \cos(\bs k' \bs r) (2\nu_{\bs k'} + 1),
\end{align}
where we use explicitly the symmetry $\bs k \to -\bs k$: the prime indicates that now the sum is performed over half of the first Brillouin zone.
The condition for ${\cal F}$ to be minimal reduces to
\begin{align}
0&=\frac{\partial {\cal F}}{\partial w_{\bs k}}=\frac{\partial {\cal F}}{\partial \nu_{\bs k}}\cr
 &=
\sum_{\mu=0}^3 \left[\frac{\partial E}{\partial F_\mu} \cos(\bs k \bs \tau_\mu) \cosh (2\theta_{\bs k}) +
                     \frac{\partial E}{\partial G_\mu} \cos(\bs k \bs \tau_\mu) \sinh (2\theta_{\bs k})\right]\cr
                     &\ \ \ \  - w_{\bs k} + \mu \cosh (2\theta_{\bs k}),\label{c1}\\
0&=\frac{\partial {\cal F}}{\partial \theta_{\bs k}}=\frac{\partial {\cal F}}{2\partial \theta_{\bs k}}\cr
 &= 
\sum_{\mu=0}^3 \left[\frac{\partial E}{\partial F_\mu} \cos(\bs k \bs \tau_\mu) \sinh (2\theta_{\bs k}) +
                     \frac{\partial E}{\partial G_\mu} \cos(\bs k \bs \tau_\mu) \cosh (2\theta_{\bs k})\right]\cr 
                     &\ \ \ \  +\mu \sinh (2\theta_{\bs k}).\label{c2}
\end{align}
Here, $\bs \tau_0 = (0,0)$ while $\bs \tau_J$, $J=1,2,3$, have been introduced above.

The condition \eqref{c2} is always equivalent to 
\begin{equation}
\tanh(2\theta_{\bs k})= \frac {A_{\bs k}}{B_{\bs k}}, \label{eq:thetak}
\end{equation}
and the condition \eqref{c1} to
\begin{equation}
w_{\bs k}=\sqrt{B_{\bs k}^2 - A_{\bs k}^2}, \label{eq:wk}
\end{equation}
where
\begin{align}
A_{\bs k}& \equiv -\sum_{\mu=0}^3 \cos(\bs k \bs \tau_\mu)\frac {\partial E}{\partial G_\mu}\cr
         & = \frac 12 \sum_J c_J \left(G_J -F_J + \cos(\bs k \bs \tau_J) \left( 1+ 2S -2 F_0 +G_0 - 4 G_J\right)\right) \cr
         & = \frac 12 \sum_J c_J \left(G_J -F_J + \cos(\bs k \bs \tau_J) \left(2S(1-|\sin \rho|) +G_0 - 4 G_J\right)\right)  ,\cr
B_{\bs k}&\equiv \mu + \sum_{\mu=0}^3 \cos(\bs k \bs \tau_\mu)\frac {\partial E}{\partial F_\mu}\cr
         & = \mu +  \sum_J c_J \left( -2 S -1 + 2 F_0 + G_J -F_J \right.\cr 
         &\ \ \ \ \ \ \left.+ \cos(\bs k \bs \tau_J)\left(S +\frac 12 + \frac 12 G_0 - F_0 + 2F_J      \right)\right)\cr
         & =\mu +  \sum_J c_J \left( -2 S(1-|\sin \rho|) + G_J -F_J \right.\cr 
         &\ \ \ \ \ \ \left.+ \cos(\bs k \bs \tau_J)\left(S(1-|\sin \rho|)+ \frac 12 G_0 + 2F_J      \right)\right), \label{eq:AkBk}
\end{align}
in the second lines of the expression for $A_{\bs k}$ and $B_{\bs k}$ we impose the generalized Takahashi constraint.

Thus, one is getting the same result as for diagonalization of quartic Hamiltonian that in momentum space is real and symmetric under 
$\bs k \leftrightarrow -\bs k$. This can be the case when the expectation value $E$ is real, but not otherwise.

At the formal level, we can use \eqref{eq:thetak} and \eqref{eq:wk} that imply
\begin{align}
\cosh (2\theta_{\bs k'}) &= \sqrt{\frac {B_{\bs k'}^2}{B_{\bs k'}^2-A_{\bs k}^2}},\cr
\sinh (2\theta_{\bs k'}) &=\frac {A_{\bs k'}}{B_{\bs k'}}\sqrt{\frac {B_{\bs k'}^2}{B_{\bs k'}^2-A_{\bs k'}^2}},
\end{align}
to write an implicit equation for the correlation functions 
\begin{align}
F(\bs r) &= \frac 1N \sum_{\bs k'} \sqrt{\frac {B_{\bs k'}^2}{B_{\bs k'}^2-A_{\bs k}^2}}\,\cos(\bs k' \bs r) (2\nu_{\bs k'} + 1), \cr
G(\bs r) &=  \frac 1N \sum_{\bs k'} \frac {A_{\bs k'}}{B_{\bs k'}}\sqrt{\frac {B_{\bs k'}^2}{B_{\bs k'}^2-A_{\bs k'}^2}}\, \cos(\bs k' \bs r) (2\nu_{\bs k'} + 1). 
\end{align}  

The following physical considerations are in order. In the zero temperature limit we are interested in, the gas of Bogoliubov excitations is expected to condense.  
Such condensation is consistent with the spin ordering 
only if the zero mode condenses, as such condensation translate into infinite range correlation in the original atomic system. 
The requirement of  zero mode to become macroscopically occupied at low temperature,  $M_0=\int_{|\bs k|<\epsilon} \nu_k \sim N n$,
implies that $w_{\bs k =0} \to 0$, which also corresponds to $|\theta_{\bs k=0}|\to \infty$. Thus, this condition can be realized only for 
$B_{\bs k =0} \sim A_{\bs k = 0}$, which implies that in the phase we are interested in, the chemical potential has to be set to zero, $\mu=0$. 
Note that this also means the occupation of each mode $\nu_{\bs k}$ is much smaller than $\frac 12$ (at least for $S=\frac 12$).     
Thus, by singling out the the zero mode and using $\nu_{\bs k}+\frac 12 \sim \frac 12$ in the expression for correlation functions, they become 
\begin{align}
F(\bs r) &\sim M_0 + \frac 1N \sum_{\bs k'\neq 0} \cosh (2\theta_{\bs k'})\,\cos(\bs k' \bs r), \cr
G(\bs r) &\sim M_0 + \frac 1N \sum_{\bs k'\neq 0} \sinh (2\theta_{\bs k'})\,\cos(\bs k' \bs r), \label{eq:FmuGmu}
\end{align}  
and the constraint \eqref{eq:gentaka} reads
\begin{equation}
M_0 + \frac 1N \sum_{\bs k'\neq 0} \cosh (2\theta_{\bs k'})\,= S  |\sin \rho| + \frac 12. \label{eq:gentakat0}
\end{equation}  

After having singled out the zero mode and constrained the occupation the function $A_{\bs k}$ and $B_{\bs k}$ should be redefined in form accordingly.
In fact only $B_{\bs k}$ gets redefined.
Indeed, by recalculating the consistency condition for an extremum of the ${\cal F}$ for the new definition of the correlation functions 
--that to say taking into account the dependence of $M_0$ on $\nu_{\bs k}$ and $\theta_{\bs k}$, as well as $\mu=0$-- 
we have
\begin{align}
0&=\frac{\partial {\cal F}}{\partial w_{\bs k}}=\frac{\partial {\cal F}}{\partial \nu_{\bs k}}
\cr
 &=
\sum_{\mu=0}^3 \left[\frac{\partial E}{\partial F_\mu} \left(\cos(\bs k \bs \tau_\mu) - 1\right) \cosh (2\theta_{\bs k}) \right.\cr
& \ \ \ \ \ \  \left.+
                     \frac{\partial E}{\partial G_\mu} \left(\cos(\bs k \bs \tau_\mu) \sinh (2\theta_{\bs k}) - \cosh (2\theta_{\bs k})\right)\right] - w_{\bs k} ,\label{c1new}\\
0&=\frac{\partial {\cal F}}{\partial \theta_{\bs k}}=\frac{\partial {\cal F}}{2\partial \theta_{\bs k}}\cr
 &=
\sum_{\mu=0}^3 \left[\frac{\partial E}{\partial F_\mu} \left(\cos(\bs k \bs \tau_\mu) - 1\right) \sinh (2\theta_{\bs k}) \right.\cr
& \ \ \ \ \ \  \left. +
                     \frac{\partial E}{\partial G_\mu} \left(\cos(\bs k \bs \tau_\mu) \cosh (2\theta_{\bs k}) -  \sinh (2\theta_{\bs k})\right)\right] .\label{c2new}
\end{align}

The above equations again imply
\begin{align*}
& \tanh(2\theta_{\bs k})= \frac {A_{\bs k}}{B_{\bs k}},\cr
& w_{\bs k} =\sqrt{B_{\bs k}^2 - A_{\bs k}^2},
\end{align*}
or alternatively
\begin{align}
\cosh (2\theta_{\bs k'}) &= \sqrt{\frac {B_{\bs k'}^2}{B_{\bs k'}^2-A_{\bs k}^2}},\cr
\sinh (2\theta_{\bs k'}) &=\frac {A_{\bs k'}}{B_{\bs k'}}\sqrt{\frac {B_{\bs k'}^2}{B_{\bs k'}^2-A_{\bs k'}^2}}.
\end{align}
The expression for $A_{\bs k}$ remains the same as in \eqref{eq:AkBk}, 
\begin{equation}
A_{\bs k}= -\sum_{\mu=0}^3 \cos(\bs k \bs \tau_\mu)\frac {\partial E}{\partial G_\mu}, \label{eq:Aknew}
\end{equation}
while $B_{\bs k}$ becomes
\begin{equation}
 B_{\bs k}= \sum_{\mu=0}^3 \left(\frac {\partial E}{\partial F_\mu} \left(\cos(\bs k \bs \tau_\mu) - 1\right) - \frac {\partial E}{\partial G_\mu}\right). \label{eq:Bknew}
\end{equation}

It is easy to check that the classical order is recovered in the limit of $S$ large.
At leading order, the minimum of the free energy is just determined by the minimum of $C$: the $\bs Q$-order found is the classical result, 
$\bs Q_{Cl}=(2\arccos(-t/2),0)$, which corresponds to $(c_1,c_2,c_3)=(\frac{ t^2-2}2, -\frac {t^2}2, -\frac {t^2}2)$.
At the next order in $\frac 1S$, which corresponds to the linear spin wave (LSW) calculation, we recover the ordinary spin-wave result: 
\begin{align}
A_{\bs k} &\to S\sum_J c_J \cos( \bs k \bs \tau_J),\cr
B_{\bs k} &\to S\sum_J c_J \left(\cos( \bs k \bs \tau_J) - 2\right), \label{eq:AkBkseed}
\end{align}
which imply 
$$w_{\bs k}=2S \sqrt{C \left(C - \sum_J c_J \cos( \bs k \bs \tau_J) \right)},$$ 
in particular 
$w_{\bs k =0}= 0$ as expected. It is easy to check that, for the classical order $\bs Q_{Cl}$, 
$w_{\bs k}$ is always real and that is by construction an extreme. In fact, as it can be checked numerically
that it is also the minimal energy solution also then the terms in $\frac 1S$, which corresponds to the case in which quadratic fluctuations are included. 
Taking into account all the terms in \eqref{eq:Ebg}, which include also $\frac 1{S^2}$ corrections
and is known as modified spin wave (MSW) approach,
the minimum condition is no longer algebraic. As in \cite{Hauke2010}, we will search for solutions
recursively, starting from the ordinary spin wave solution above. 
The absence of a pronounced minimal value will signal the existence of possible spin-liquid phase. 
In order to find the optimal $\bs Q=(Q_x,Q_y)$, we have to impose that the gradient is zero
\begin{align}
0&=\frac{\partial{\cal F}}{\partial Q_x}= \sum_J \frac{\partial E}{\partial c_J} \frac{\partial c_J}{\partial Q_x},\cr
0&=\frac{\partial{\cal F}}{\partial Q_y}= \sum_J \frac{\partial E}{\partial c_J} \frac{\partial c_J}{\partial Q_y}.
\end{align}

\section{Result from the modified spin wave analysis}
In the previous section, we have derived a modified spin wave theory for the XX spin model on a triangular lattice. 
We will now extract concrete results from this theoretical framework. This amounts for a minimization problem of the free energy, 
which is complicated due to the large amount of variables. Using the procedure described in the subsection below, 
we manage to perform minimization even for large lattices with hundreds of sites. As the result, we then obtain the phase diagram for 
a realistic experimental system as a function of the hopping anisotropy $t$.

\subsection{Optimization and stability}

In search for a long-range order in the quartic case, we adopt an iterative procedure.
We start from the ordinary spin-wave \eqref{eq:AkBkseed} solution with $\bs Q = \bs Q_{Cl}$ as 
initial configuration. The recursive procedure works as follows. 
First, the values of $A_{\bs k}$, $B_{\bs k}$ at the cycle $m$ are 
used to get the new correlation functions $F_\mu, G_\mu$, using 
Eq.  \eqref{eq:FmuGmu}. Once the correlations are substituted in the free energy, 
which at zero temperature reduces to the expectation value of the energy \eqref{eq:Ebg}, 
the latter becomes a function of the ordering vector $\bs Q$ only, $E=E(\bs Q)$. 
The new value at the cycle $m$  of $\bs Q$ is, thus, determined by minimizing the $E(\bs Q)$ in the neighborhood of 
optimal value of $\bs Q$ at the cycle $m$. Finally, \eqref{eq:Aknew} 
and \eqref{eq:Bknew} are used to update $A_{\bs k}$ and $B_{\bs k}$ as a function of the correlation functions and of the order vector.
Convergence of the iterative process is assumed when the difference 
between the old and the updated values of $A_{\bs k}$ and $B_{\bs k}$ are below a certain threshold.

\begin{figure}
\includegraphics[width=0.99\columnwidth]{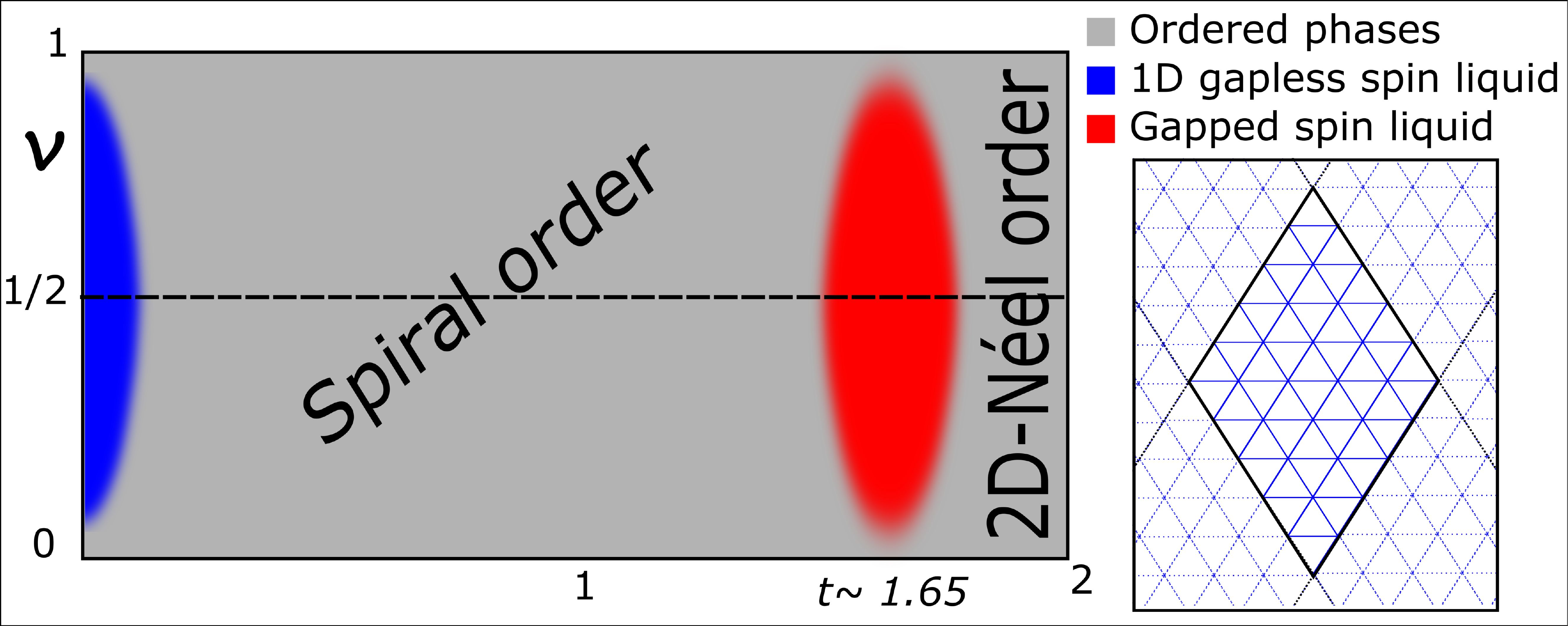}
\caption{\label{fig:rhombic}
(Color online) Expected phase diagram as function of the filling within the MSW approach. The two ordered phases are spiral order, $0.2\lesssim t \lesssim 1.55$, and
2D-N\'eel order. Inset: a $6 \times 6$ lattice of rhombic shape with periodic boundary conditions. }
\end{figure}

We have benchmarked the performance of this iterative approach against brute force minimization of the energy 
as function of the free parameters $\theta_{\bs k}$ and $\bs Q$ for different shapes and sizes
of lattices with periodic boundary conditions. While the success and efficiency of the iterative approach, 
i.e. the number of iterations needed for achieving convergence, strongly depends on the shape of the lattice, it performs
generally better than a brute force minimization and the scalability with the lattice size is pretty good. 
Best performance is achieved for rhombic lattices, see Figure \ref{fig:rhombic}.  
Converge or failure occurs after few tens of iterations. The latter manifests when $|A_{\bs k}|$ becomes greater than $|B_{\bs k}|$ 
for some ${\bs k}$, which corresponds to $w_{\bs k}$ becoming imaginary. In fact, more than a real instability, 
the absence of convergence signals that the approximation we have used of neglecting the occupation $\nu_{\bs k}$ of the modes $\bs k \neq 0$ is not respected. 
That is to say, the physical assumption of the existence of an ordered phase behind the spin-wave analysis is not verified. 
The comparison between the iterative approach and the brute force minimization of the free energy, which we have performed   
without assuming $\nu_{\bs k}\ll 1$  on $L\times L$ rhombic lattices with $L$ up to 10, confirmed this scenario.

Next we have extended our iterative minimization to larger lattices. We have first studied the half-filling case for $L=24, 100$ and for the infinite $L$ limit, 
obtained by replacing the sum over $\bs k$ with an (numeric) integral 
over the first Brillouin zone. 

\subsection{Phase diagram predicted by spin wave at half-filling}

\begin{figure}
\includegraphics[width=0.98\columnwidth]{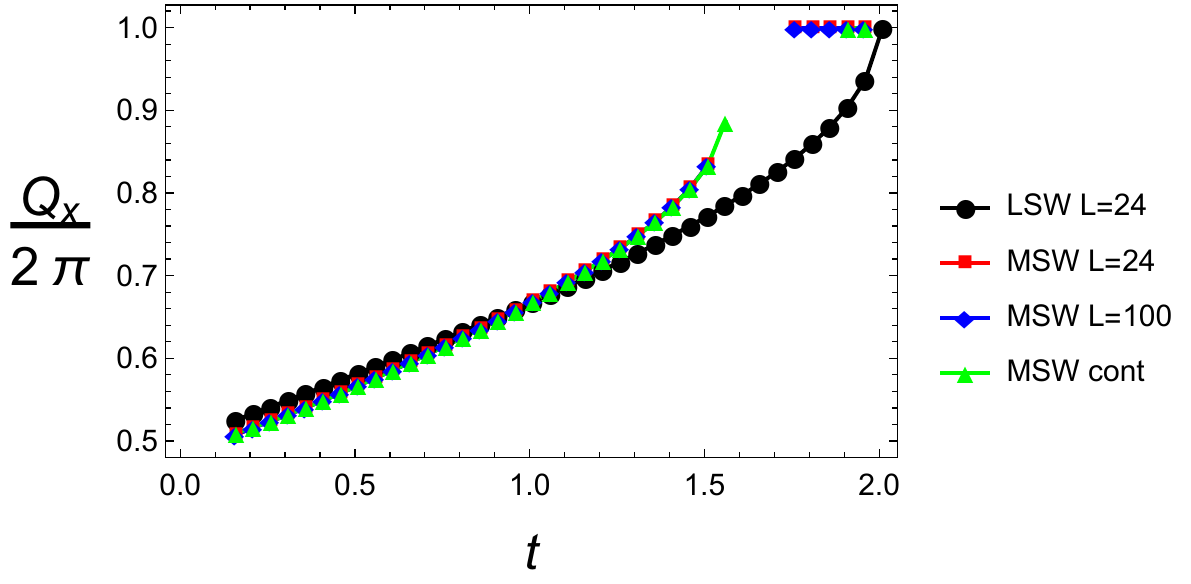}
\caption{\label{fig:qxhalf}
(Color online) Values of the optimal $Q_x$: comparison of results from LSW and from MSW for different sizes of the rhombic-shape lattices with periodic boundary conditions. }
\end{figure}

\begin{figure}
\includegraphics[width=0.98\columnwidth]{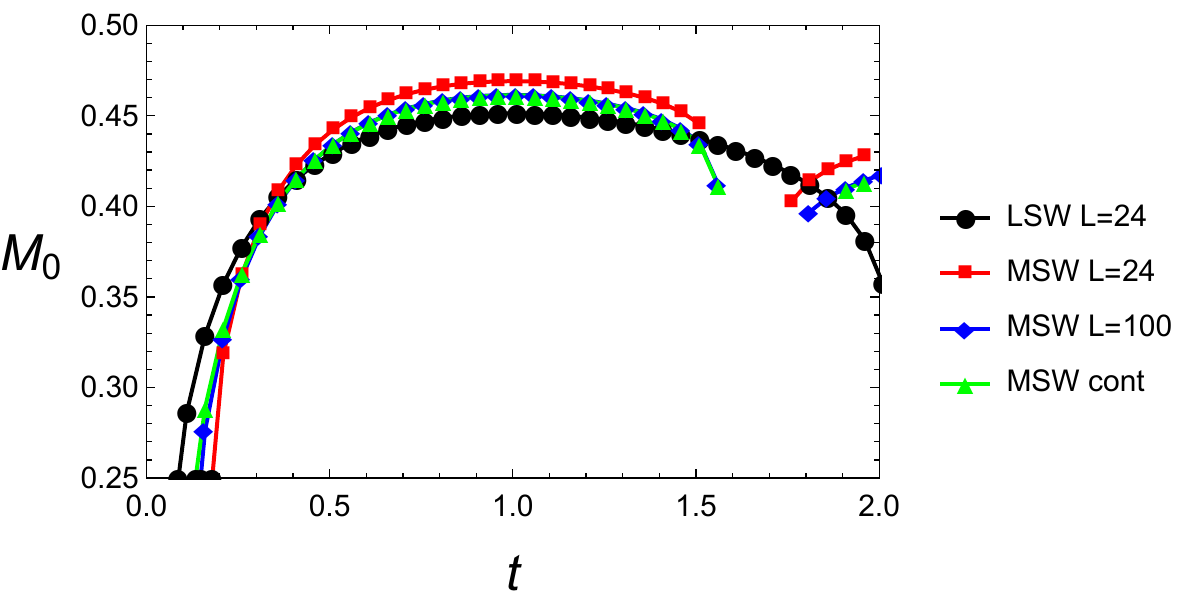}
\caption{\label{fig:M0half}
(Color online) Occupation of the ground state at zero momentum corresponding to the ordered solution: 
comparison of results from LSW and from MSW for different sizes of the rhombic-shape lattices with periodic boundary conditions.}
\end{figure}

\begin{figure}
\includegraphics[width=0.98\columnwidth]{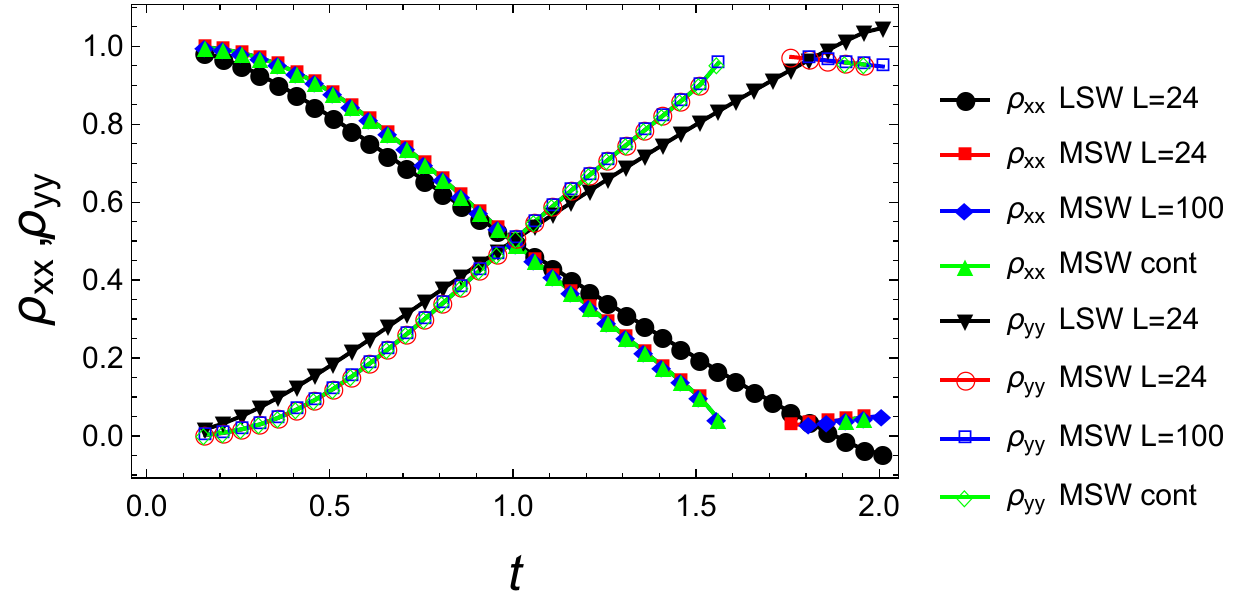}
\caption{\label{fig:rhoxxyyhalf}
(Color online) Values of the rescaled susceptibility $\rho_{xx}$ and $\rho_{yy}$:
 comparison of results from LSW and from MSW with different lattice
  dimensions. Around the non-convergence window $\rho_{xx}$ is small,
   signaling instability of the order. Around the non-convergence window $\rho_{yy}$ is 
   large as expected because the $y$-component of the order vector $Q_y$ is the same for spiral and 2D-N\'eel order. 
Instead, $\rho_{yy}$ signals the instability that leads to 1D-N\'eel order for $t\sim 0$. }
\end{figure}

We have first started by studying the half filling case, $\rho =\frac {\pi}2$. Our results are very close to the one of \cite{Hauke2010} and display the same qualitative behavior (see Figure \ref{fig:rhombic}). 
In particular, we observe a failure of convergence around $t\sim 0$ and for  $t$ between $1.55 - 1.8$. The first region is easily explained: in the limit $t\to 0$ the system reduces to disconnected 
$1D$-XX chains that can order separately in 1D-N\'eel orders with arbitrary relative phases. Thus, there is a huge degeneracy in the groundstate that should correspond to a gapless spin-liquid phase. 
The region around $t\sim 1.65$ appears
at the interface between two classically ordered phases, the spiral order and a 2D-N\'eel order, which appear at lower and higher values of $t$, respectively. Both phases are well described by 
the classical order ansatz we used. It is worth noticing that the initial condition and the reflection symmetry of the Hamiltonian around the $x$-axis implies that
our solution is respecting such symmetry i.e.  the ordering vector remains parallel to the $x$-axis and the correlation functions
respect the relations $F_2=F(\bs \tau_2)=F(\bs \tau_3)=F_3$, $G_2=G(\bs \tau_2)=G(\bs \tau_3)=G_3$. 
This implies that we can work at fixed $Q_y= 0$. For this choice, the 2D-N\'eel order corresponds to 
$Q_x=2 \pi $, while the spiral order corresponds to $Q_x$ smoothly interpolating between $2\pi$ and $\pi$ for decreasing values of the anisotropy $t$. 
While at the classical level, the 2D-N\'eel order is predicted to be stable for $t\ge 2$, the quantum corrections incorporated by MSW approach stabilize 
it also for lower values of $t$, as displayed in Figure \ref{fig:qxhalf}. Similar results are obtained by exact diagonalization, see Figure \ref{fig:sftrap}(a). 
By reducing further the values of $t$ the system enters in a non-ordered phase signaled by the absence of points from MSW. While in the neighboring regions above and below the no-convergence 
window the occupancy of the zero-momentum states remains large, see Figure \ref{fig:M0half}, the values of the relative susceptibility $\rho_{xx}$ is small in the vicinity of such window, Figure \ref{fig:rhoxxyyhalf}.
Similarly to \cite{Hauke2010}, we estimate the susceptibility by calculating the Hessian of the energy for fixed correlation functions at the minimum. 
In order to get an adimensional quantity we divide by the absolute value of the energy minimum, thus, $\rho_{xx} = \frac 1E \frac{\partial^2 E}{\partial Q_x^2}$, and $\rho_{yy} = \frac 1E \frac{\partial^2 E}{\partial Q_y^2}$.
Note that $\rho_{xy} = \frac 1E \frac{\partial^2 E}{\partial Q_x\partial Q_y}$ is identically zero because of the symmetry argument given above. As expected $\rho_{yy}$ is not signaling any instability for $1.5\le t\le 2$ 
--the optimal $Q_y$ is identical for the spiral and 2D-N\'eel order-- while it detects the instability at $t\sim 0$, see Figure \ref{fig:rhoxxyyhalf}. 
While for all the observables represented in Figure \ref{fig:qxhalf}-\ref{fig:rhoxxyyhalf}
the MSW results deviate considerably from the ones predicted by the LSW, they are  
quickly converging to a stable behavior for moderate size-lattices -- for a rhombic shape lattice $L\times L$ the deviation between $L=24$ and the continuous limit are tiny.

\subsection{Phase diagram predicted by spin wave at generic filling}

\begin{figure}
\includegraphics[width=0.98\columnwidth]{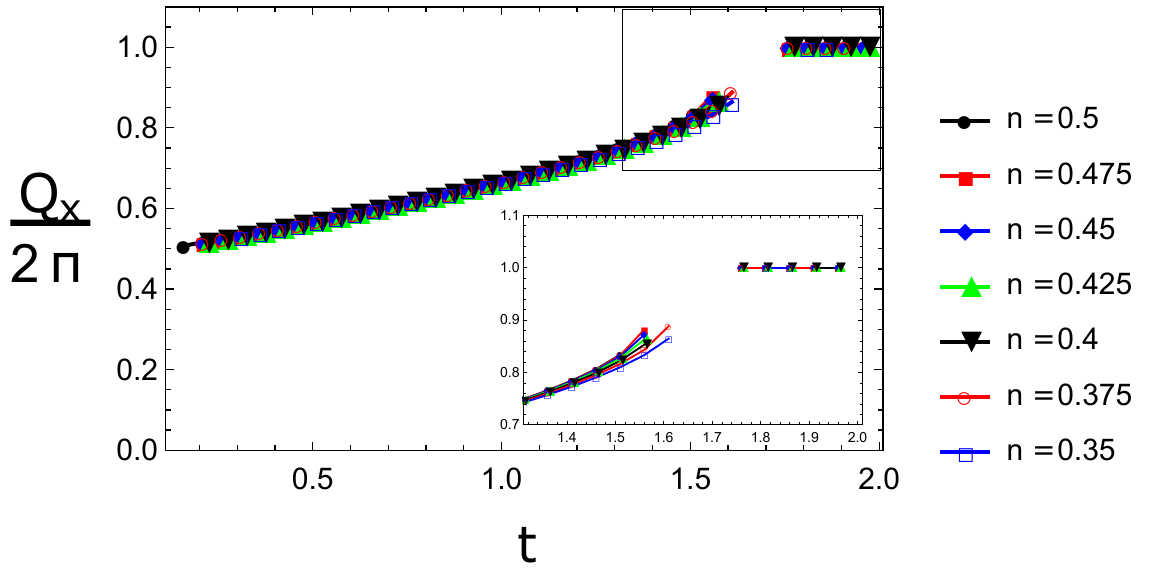}
\caption{\label{fig:qxfill}
(Color online) The value of the optimal $Q_x$ depends in sizable way on the filling only close to the transition to the non-ordered region. }
\end{figure}

\begin{figure*}
\includegraphics[width=0.98\columnwidth]{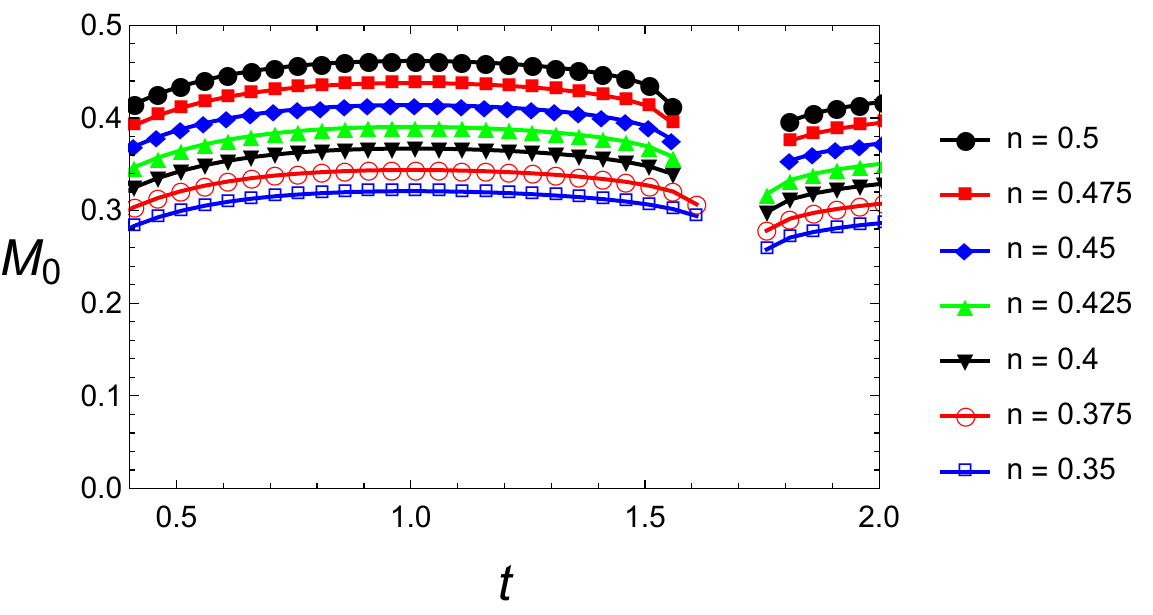}\ \ \ \   \includegraphics[width=0.98\columnwidth]{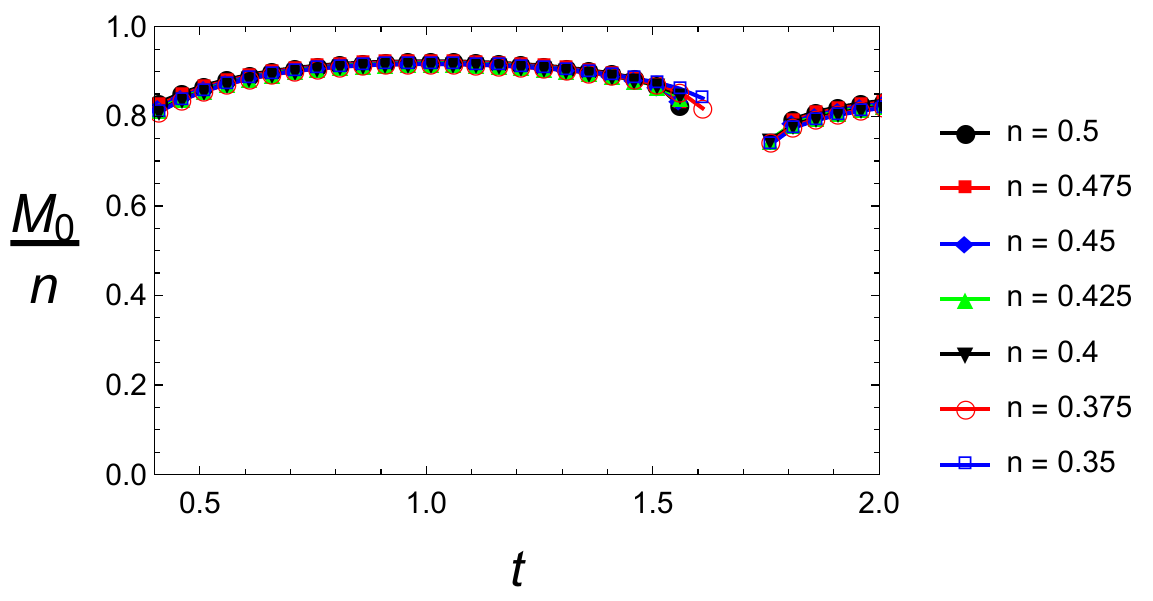}
\caption{\label{fig:M0fill}
(Color online) (a) Occupation of the the state at zero momentum for different filling: the non-ordered region shrinks smoothly with the filling. 
(b) The condensate fraction is independent of the filling in the ordered regions. }
\end{figure*}

\begin{figure}
\includegraphics[width=0.98\columnwidth]{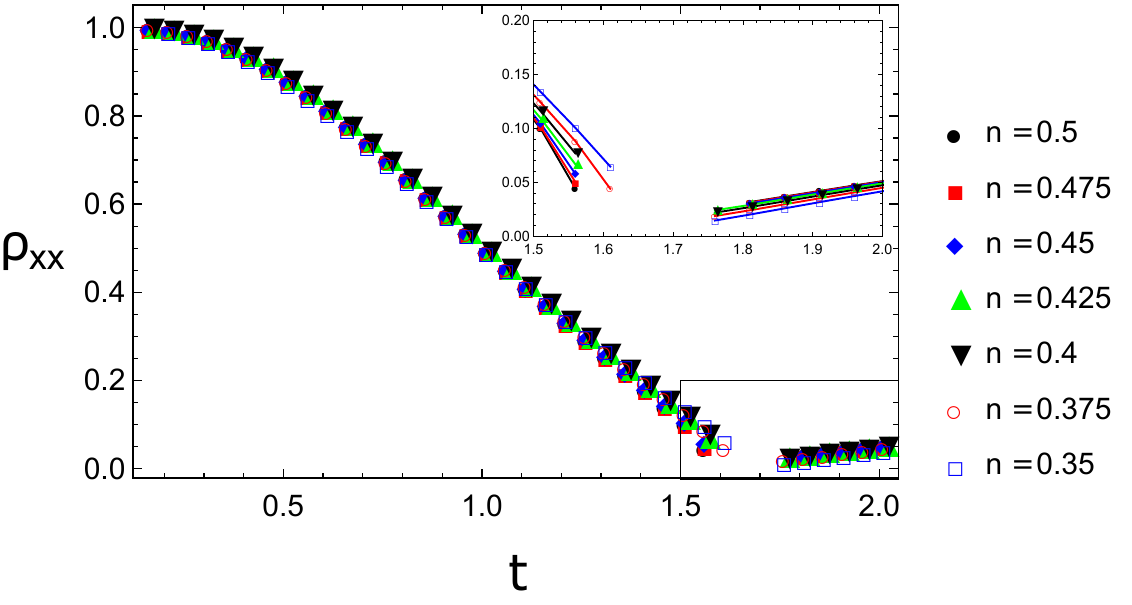}
\caption{\label{fig:rhoxxfill}
(Color online) The value of the rescaled susceptibility $\rho_{xx}$ depends sizably on the filling only close to the transition to the non-ordered region. 
A slighter increase of $\rho_{xx}$ at lower $n$ translates in an increased stability of the ordered phases. }
\end{figure}

\begin{figure}
\includegraphics[width=0.98\columnwidth]{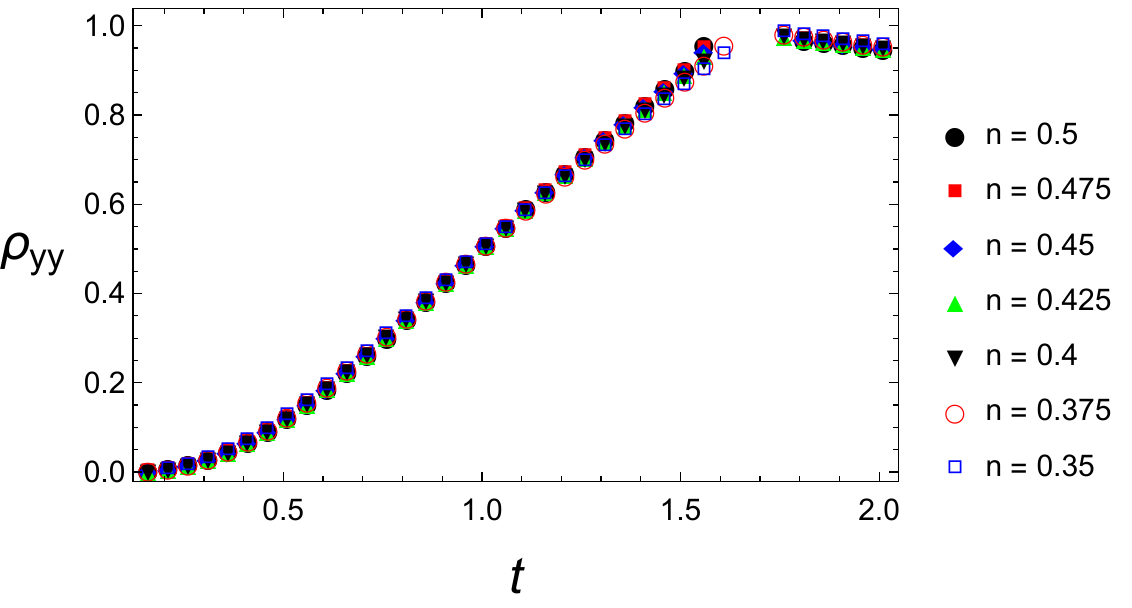}
\caption{\label{fig:rhoyyfill}
(Color online) The value of the rescaled susceptibility $\rho_{yy}$ does not depend considerably on the filling. 
Indeed, it is sensitive only on the transition at very low $t$ to the the 1D behavior, which is unaffected by the filling, while is order 1 around the transition at $t\sim 1.65$.    }
\end{figure}

Then, we have considered lower values of $\rho$ between $0$ and $\frac{\pi}2$, corresponding to lower densities of Bogoliubov excitations $n= \frac 12 \sin \rho = \frac 12 \sqrt{\nu(1-\nu)}$, where $\nu$ is the filling.
We have considered the same observables as in the half-filling case. We have found again that the results quickly saturate to a stable value for growing size of the lattices. For simplicity, we present here the results 
$L\times L$ rhombic-shaped lattices with periodic boundary conditions for $L=100$. 
First, we notice that the values of the optimal order vector $\bs Q$ remains substantially unchanged with respect to the half-filling case. 
While by construction $Q_y=0$, the $x$-component of the order vector $Q_x$ displays a
moderate dependence on $n$ only close to the non-convergence window, Figure \ref{fig:qxfill}. 
In fact, the non-convergence window changes: while it remains centered around $t\sim 1.65$, its extension shrinks smoothly while $n$ decreases.
Indications of such behavior can be detected both in the condensate fraction and in the susceptibility. 
Indeed, the shrinking of the non-convergence window is well evident 
in Figure \ref{fig:M0fill}(a) where the occupation of the zero mode $M_0$ is depicted. As expected $M_0$ is directly proportional to $n$, that is to say the condensate 
fraction $\frac{M_0}n$ depicted in Figure \ref{fig:M0fill}(b) is independent of $n$. 
This behavior supports the picture that the nature of the ordered phases is unchanged while their stability increase by moving away from half filling $n=\frac 12$. 
Further confirmation comes from the calculation of the relative susceptibilities $\rho_{xx}$ and $\rho_{yy}$. 
While $\rho_{yy}$ does not display a strong dependence on $n$, Figure \ref{fig:rhoyyfill}, $\rho_{xx}$
displays a sizable dependence on $n$ only around the non-convergence window. In particular, $\rho_{xx}$ weakly 
increases when $n$ decreases, showing that the ordered phase gets  smoothly more stable, Figure \ref{fig:rhoxxfill}.
Thus, we conclude that by moving out of half-filling the conjectured spin-liquid phase signaled by the non-convergence 
window of MSW does not disappear but shrinks rather gently.

\section{Exact diagonalization study}
In this section we will study the Hamiltonian (\ref{eq:H1}) by means of exact diagonalization. Therefore, we first note that it conserves the $z$-component of total spin, 
$S_z \equiv \frac{1}{N} \sum_i S_i^z$.  This symmetry reflects conservation of particles, and allows to work in Hilbert space blocks with fixed spin polarization. 
Using this symmetry, we are able to exactly diagonalize systems of 24 sites, as depicted in Figure \ref{fig:lattice}. 
As in the spin-wave analysis, we will first consider the system within a local density approximation, assuming homogeneity within shells of different $S_z$. 
Our exact diagonalization study is exptected to capture the system behavior in the center of the trap, and we set $V_i=0$. 
Afterwards, we study effects of the trapping potential on small scales, diagonalizing Eq. (\ref{eq:H1}) at finite $V_i$. 
The exact diagonalization study presented here covers the case at half filling ($S_z=0$) known from Ref. \cite{Schmied2008,Hauke2010}, 
with a possible quantum spin liquid for $t \approx 0.5$ and $t \approx 1.5$. We extend this study to other polarization sectors, 
which become relevant if the trap leads to an increased density in the center. 

Of course, in a real experiment the central area would be surrounded by rings with decreasing density, while the exact diagonalization study considers a scenario with hard walls.
 We will therefore, in Section \ref{DMRG}, use DMRG methods to demonstrate that the hard wall assumption becomes reasonable for sufficiently steep trapping potentials and/or low densities.

\subsection{Homogeneous system ($V_i=0$)}
\begin{figure*}
\includegraphics[width=0.95\columnwidth]{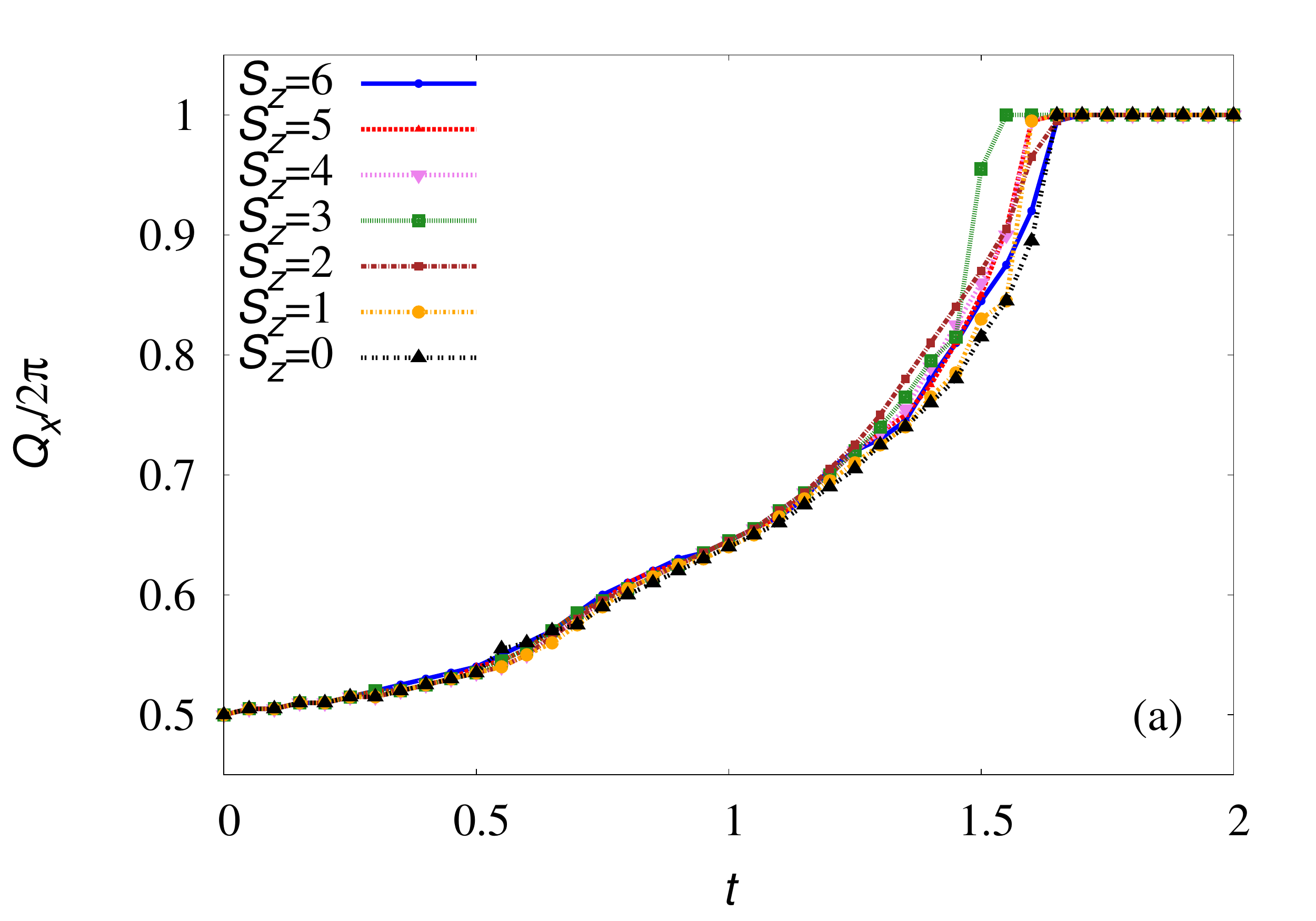}
\includegraphics[width=0.95\columnwidth]{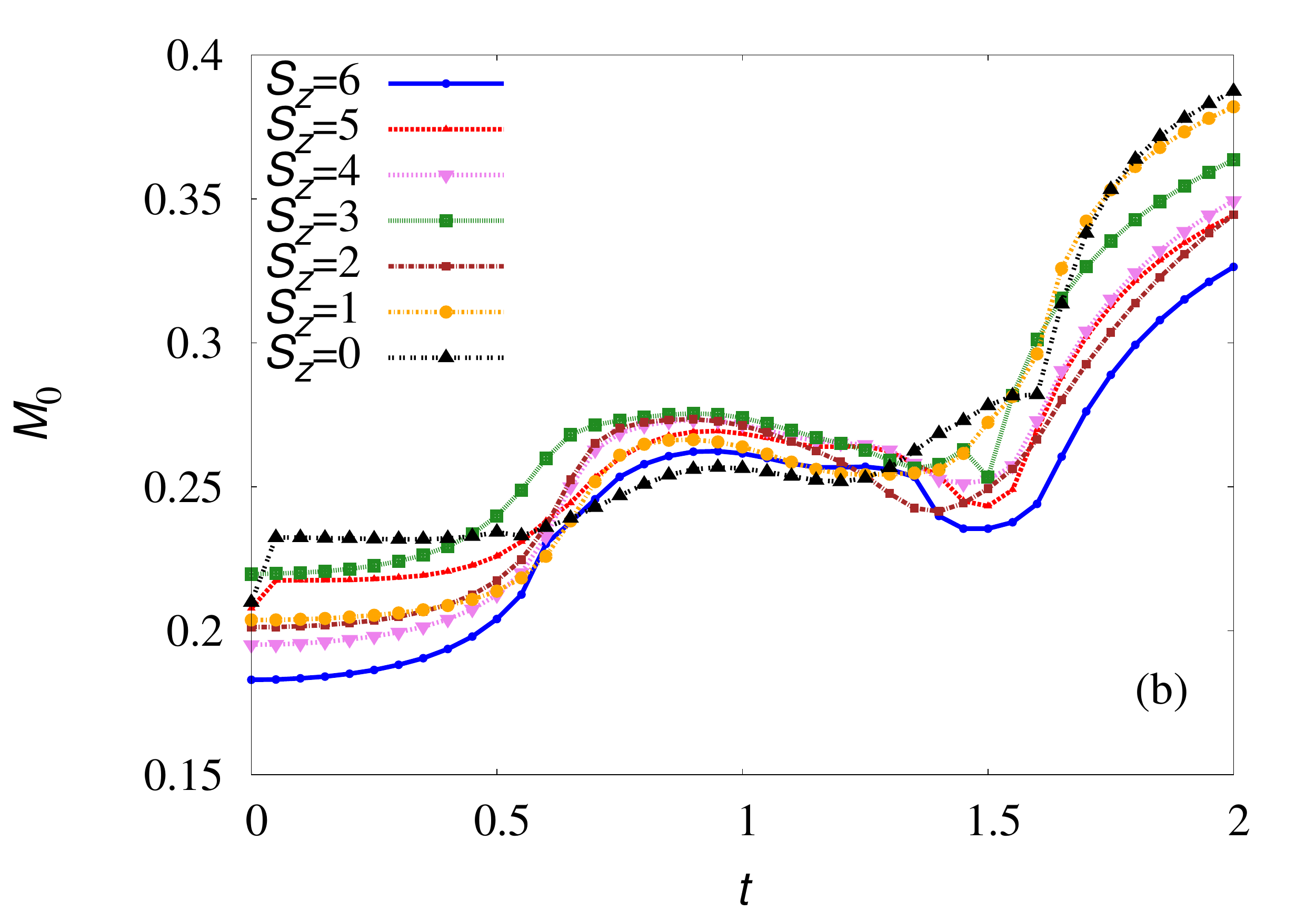}
\caption{\label{fig:sf} (Color online)(a) Position of peak of structure factor $S({\bf q})$ as a function of anisotropy $t$ for a homogeneous system ($B_i=0$) at different $S_z$. 
(b) Order parameter $M_0$ as defined in Eq. (\ref{M0}).}
\end{figure*}

As an experimentally accessible quantity which allows to chararacterize the order of the system, we have calculated the structure factor $S({\bf q})$:
\begin{align}
 S({\bf q}) = \frac{1}{N} \sum_{i \neq j} \exp\left[ i {\bf q} \cdot ({\bf r}_i - {\bf r}_j) \right] \langle S_i^+ S_j^- + {\rm h.c.} \rangle.
\end{align}
Here $\langle \cdot \rangle$ denotes the quantum average of the ground state. The existence of a pronounced peak signals an ordered phase, 
and the momentum space position ${\bf Q}$ of the peak further characterizes this order. As an order parameter $M_0$ we define
\begin{align}
\label{M0}
 M_0 = \sqrt{S({\bf Q})/N}.
\end{align}
If we don't restrict ourselves to the first Brillouin zone (having a hexagonal geometry), we can, at all fillings and for all anisotropies, find a global maximum of the structure factor for $Q_y=0$. 
In Figure \ref{fig:sf}(a), we have plotted the corresponding $x$-component of ${\bf Q}$, and $M_0$ as a function of $t$ in different polarization sectors. 
Remarkably, the peak position hardly depends on the spin polarization for most values of $t$, 
except for a small region around $t \approx 1.5$, where ${\rm d}Q_x/{\rm d}t$ tends to infinity. 
This means that a trapped system, composed of subsystems with different $S_z$, should exhibit a similar structure factor as the homogeneous system, 
except for a possible broadening of the peak near $t \approx 1.5$. For $S_z=0$, the two limiting cases ${\bf Q}=\pi \hat x$ and ${\bf Q}=2\pi \hat x$, 
reached for $t=0$ and $t > 1.5$, correspond to an intrachain N{\'e}el order, and to a square-lattice N{\'e}el order, respectively.

In Figure \ref{fig:sf}(b), we show the order parameter $M_0$ obtained from the structure factor in different polarization sectors. In contrast to $Q_x$, 
the different curves of $M_0$ do not overlap, although again many qualitative properties are shared between different polarization sectors: 
For all $S_z$, the curve of $M_0$ as a function of $t$ can roughly be described in the following way: 
For $t \lesssim 0.5$, the curves are flat at a low value of $M_0$. The value of $M_0$ then quickly increases, 
before the curves become flat again for $0.75 \lesssim t \lesssim 1.25$. 
The curves then exhibit a dip or a kink around $t \approx 1.5$, before they increase again.

The occurrence of a quantum spin liquid phase, as a breakdown of the ordered phases, 
should be reflected by small values of $M_0$. The dip at $t \approx 1.5$ can therefore be taken as a signal for spin liquid behavior. 
For larger systems, such signal is also found near $t \approx 0.5$, as shown in Ref. \cite{Hauke2010} for $S_z=0$ in a 20$\times$20 lattice studied via PEPS.

Useful information about the phase of the system is contained also in the entanglement spectrum, 
obtained from the eigenvalues of the reduced density matrix $\rho_{\rm L} \equiv {\rm Tr}_{\rm R} \ket{\Psi}\bra{\Psi}$.
 Here, $\rm Tr_R$ denotes a trace over half spins, localized on the right side  of the lattice. 
 The entanglement spectrum is then defined as $\lambda_i = - \log \rho_i$, where $\rho_i$ denote the eigenvalues of $\rho_{\rm L}$. 

\begin{figure}
\includegraphics[width=0.98\columnwidth]{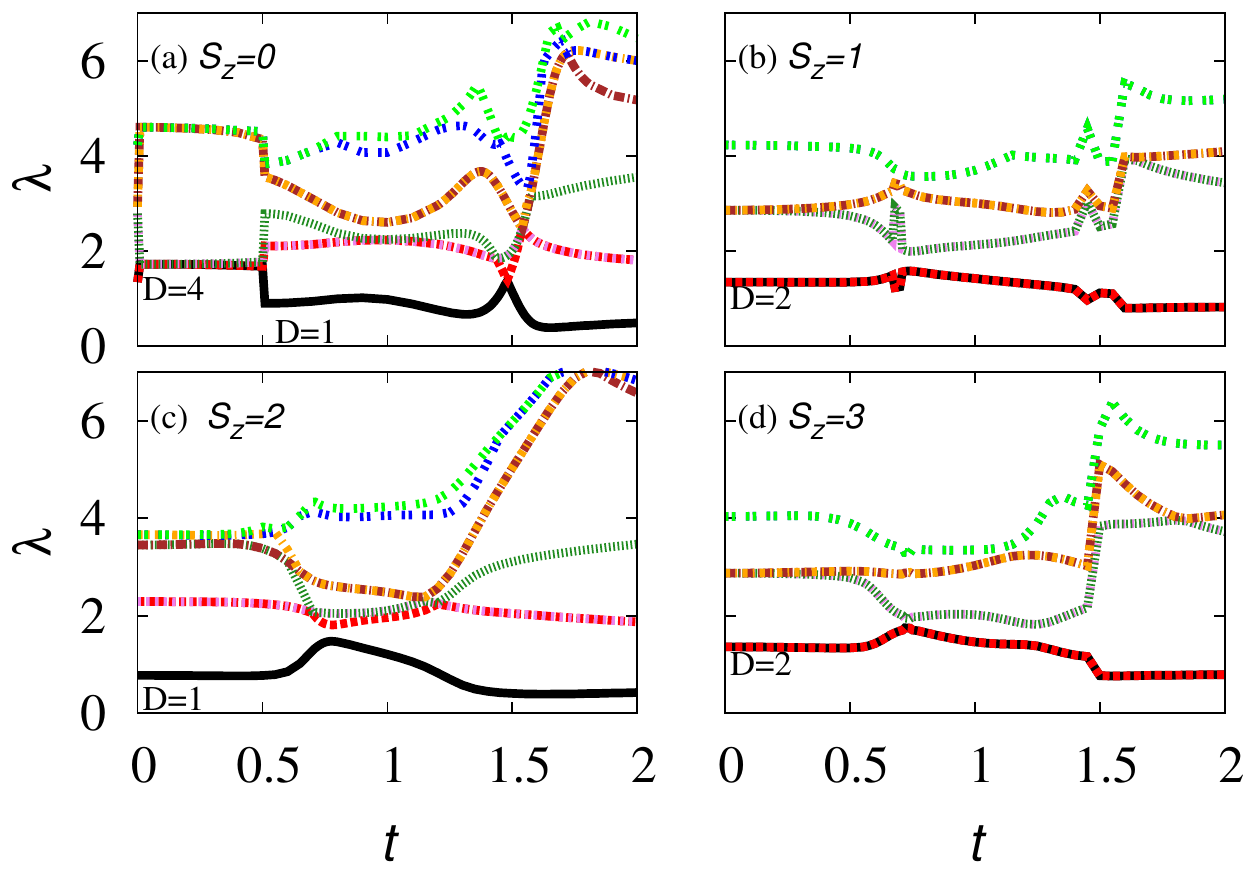}
\caption{\label{fig:es}(Color online) Entanglement spectrum (8 lowest values) for different spin polarizations. $D$ denotes the number of degenerate levels in the ground state.}
\end{figure}

In Figure \ref{fig:es} we plot the eight lowest values of the entanglement spectrum in a homogeneous system as a function of the anisotropy $t$ for different spin polarizations. 
Some qualitative features, observed in the behavior of the order parameter $M_0$, are reflected by the entanglement spectrum: all eigenvalues remain flat in the regime $0< t \lesssim 0.5$.
The curves are also relatively flat for intermediate values $0.75 \lesssim t \lesssim 1.4$. 
In contrast to these flat regimes, the curves exhibit quick or even sudden changes in the regimes $0.5 \lesssim t \lesssim 0.75$ and $1.4 \lesssim t \lesssim 1.7$: 
for $S_z=0$, the fourfold quasidegeneracy of the lowest level is abruptly lifted at $t \approx 0.5$. 
This sudden change  in the entanglement entropy indicates a second-order phase transition. For $t \approx 1.8$, 
the ground state level at $S_z=0$ exhibits a crossing, accompanied by highly non-monotonous behavior in all levels. 
For $S_z=1$, the several levels exhibit pronounced dips around $t \approx 0.69$ and  $t \approx 1.5$, 
without affecting the two-fold degeneracy of the ground state level. For $S_z=2$ and $S_z=3$, we observe avoided crossings of the ground state level near $t \approx 0.75$. 
Sudden changes of all eigenvalues occur at $t \approx 1.5$ in the $S_z=3$ sector. 

In summary, the behavior of the entanglement spectrum suggests that, independently from the spin polarization, 
changes of the ground state occur in the two regimes: for $0.5 \lesssim t \lesssim 0.75$, and for $1.4 \lesssim t \lesssim 1.7$. 

\subsection{Inhomogeneous system ($V_i \neq 0$)
}

In the previous paragraph, we have shown that the system, to some extent, behaves similarly in different polarization sectors upon tuning the anisotropy $t$. 
This allows one to argue that the same behavior should persist in a shallow trap, where the system is approximated by homogeneous subsystems of different polarizations. 
In the present paragraph, we go a step further, and analyze the effect 
of a trap on short scales by diagonalizing Hamiltonian (\ref{eq:H1}) for $V_i = \frac{m}{2} \omega^2 {\bf r}_i^2 = \eta  {\bf r}_i^2$, with $\eta = 0.1$ (in units $t_1/a^2$) 
for typical trapping frequencies of 40 Hz. 
We will focus on the $S_z=0$ sector, corresponding to half filling.

\begin{figure}
\includegraphics[width=0.98\columnwidth]{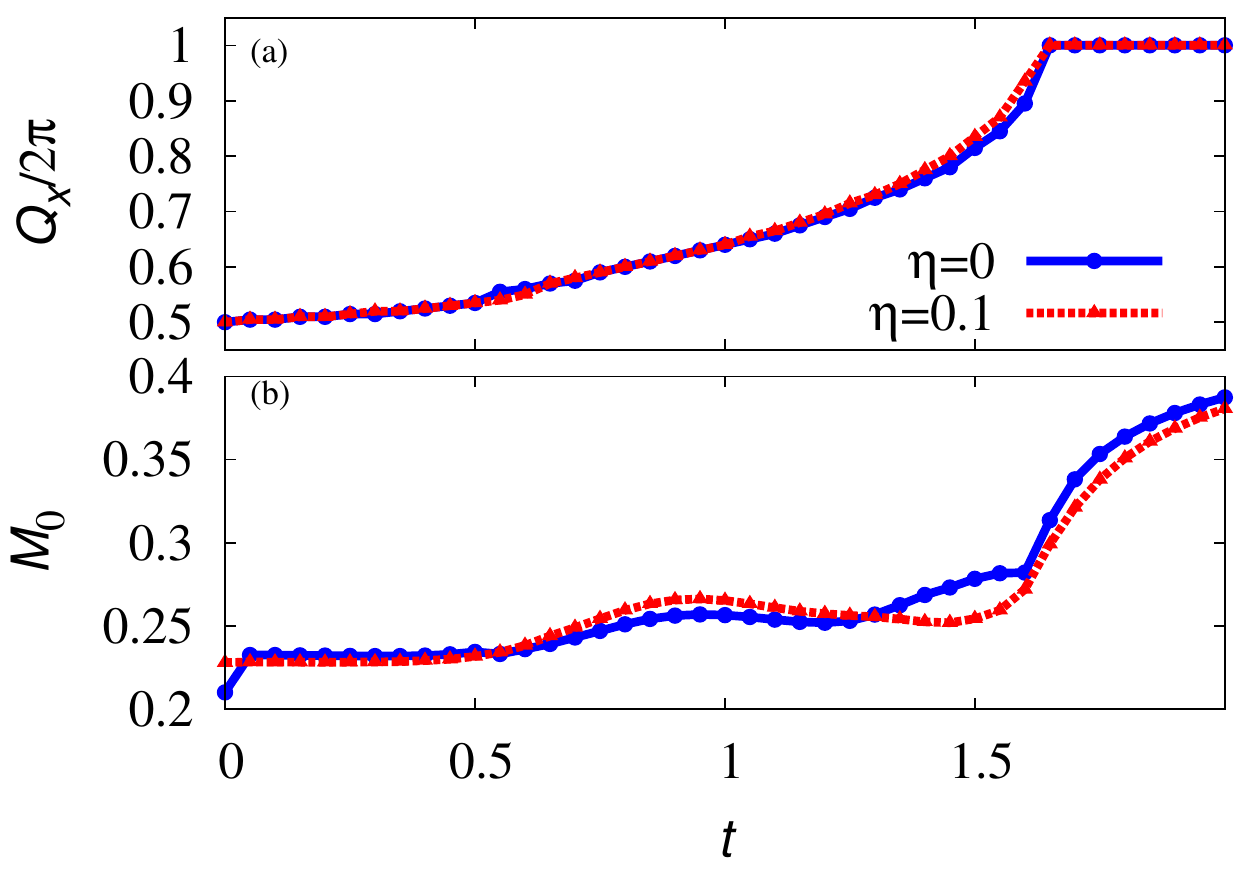}
\caption{\label{fig:sftrap}
(Color online) (a) Position of peak of structure factor $S({\bf q})$ as a function of anisotropy $t$ at $S_z=0$ in a homogeneous system and for $\eta=0.1$. 
(b) Order parameter $M_0$ as defined in Eq. (\ref{M0}).}
\end{figure}

\begin{figure}
\includegraphics[width=0.98\columnwidth]{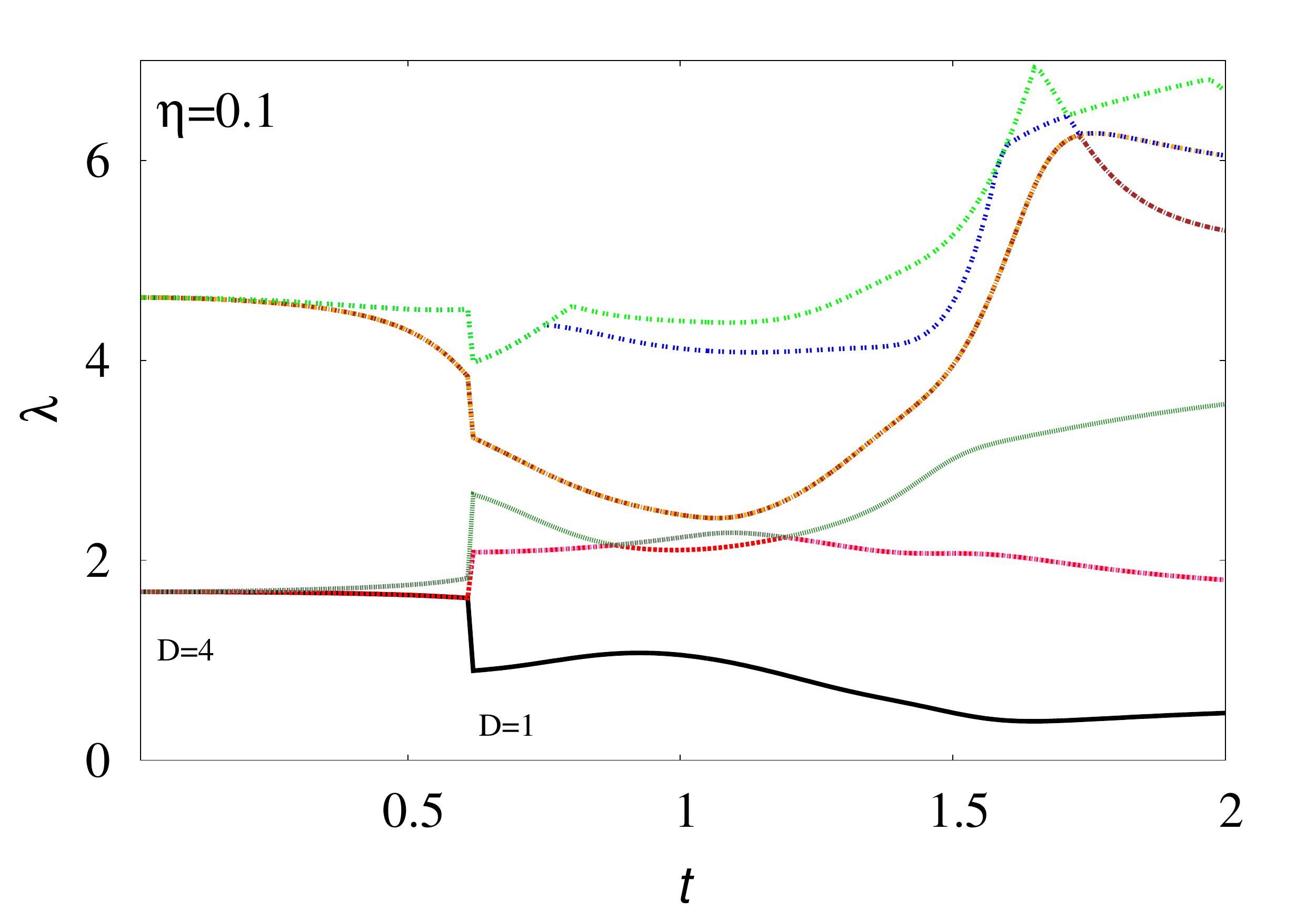}
\caption{\label{fig:estrap}
(Color online) Entanglement spectrum (8 lowest values) at $S_z=0$ in a trapped system at $\eta=0.1$. $D$ denotes the number of degenerate levels in the ground state.}
\end{figure}

On the small lattice studied here, the inhomogeneities introduced by the trap, are rather weak: For the isotropic system, $t=1$, 
we find an average population of 0.46 atoms on the 14 sites at the edge of the lattice, while the remaining 10 sites have an average population of 0.56 atoms. 
Accordingly, also the structure factor is barely modified: as shown in Figure \ref{fig:sftrap}(a), 
the peak position $Q_x$ is practically indistinguishable for the two cases $\eta=0$ and $\eta=0.1$. Also the order parameter $M_0$, 
shown in Figure \ref{fig:sftrap}(b), exhibits a similar shape, though slightly smoothened near $t = 1.5$. 
Also the entanglement spectrum, plotted in Figure \ref{fig:estrap}, shares important qualitative features with the one of the homogeneous system shown in Figure \ref{fig:es}(a): 
For small values of $t$ the ground state level has a perfect twofold degeneracy, and a fourfold quasidegeneracy. 
Again, the lifting of the degeneracy occurs abruptly near $t \approx 0.5$, although the precise value of the anisotropy is slightly increased by the trap. 
On the other hand, the level crossing observed in the homogeneous case around $t \approx 1.5$ does not take place in the trapped scenario.

\begin{figure*}
\includegraphics[width=0.6\textwidth]{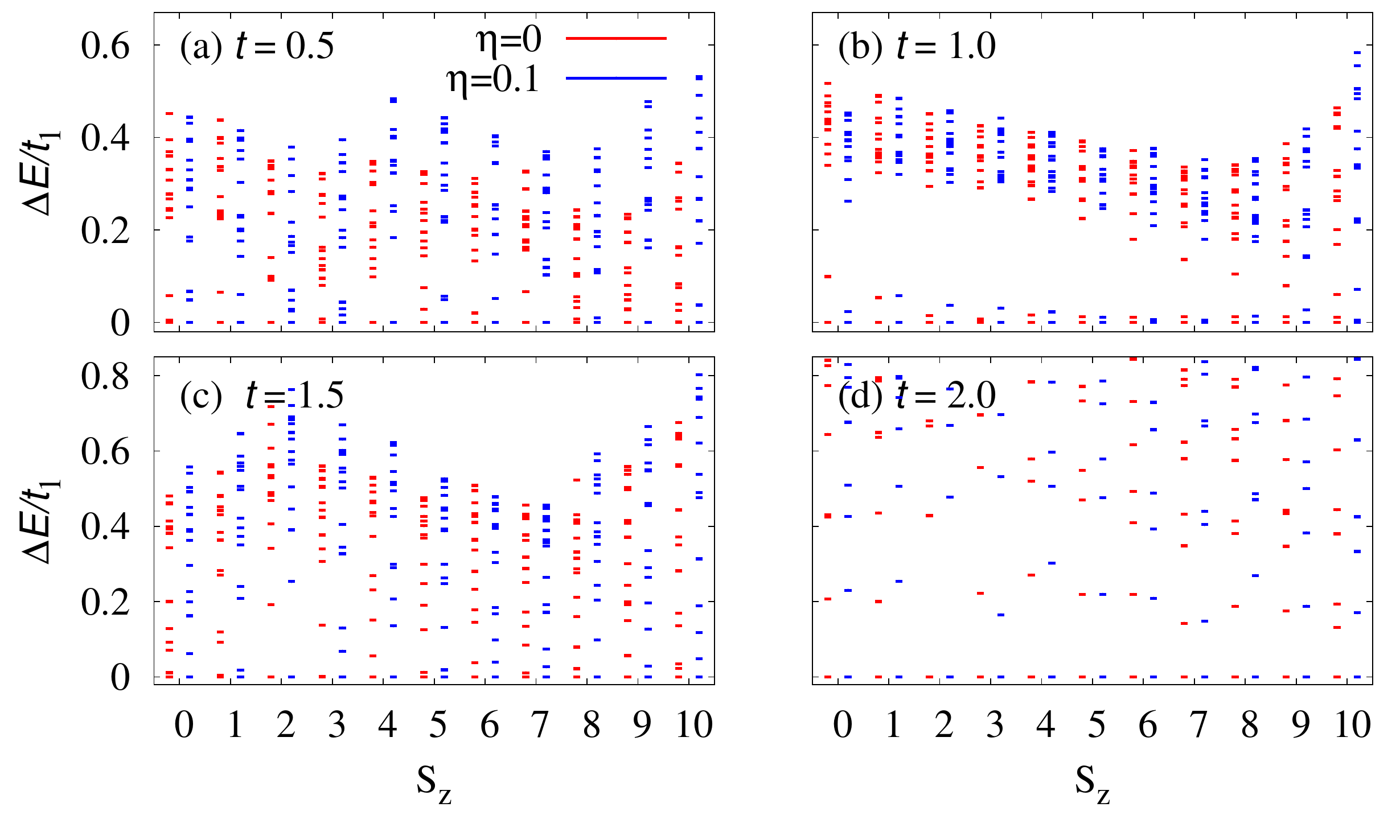}
\includegraphics[width=0.38\textwidth]{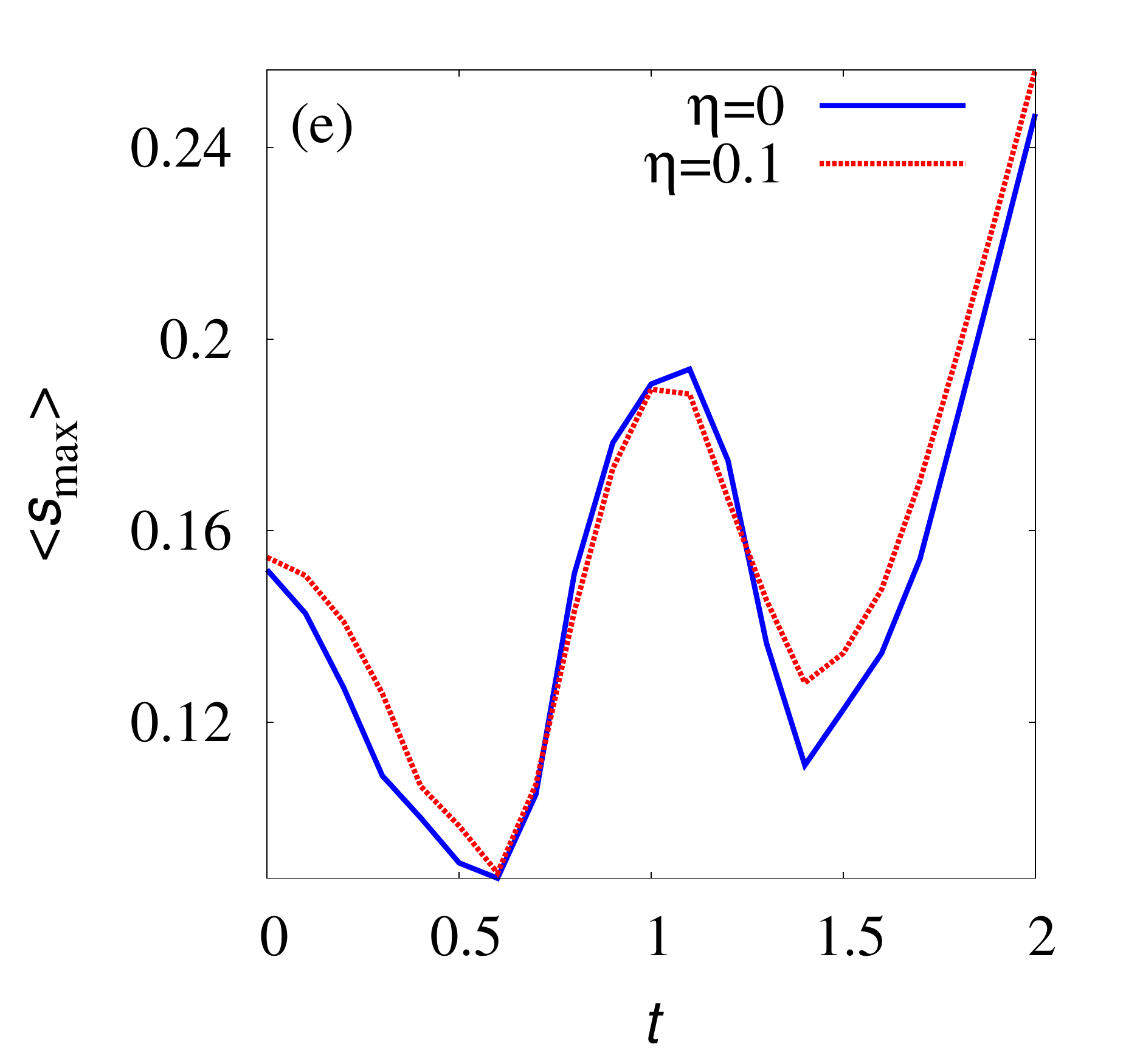}
\caption{\label{fig:spectra}
(Color online) (a--d) Energy versus spin polarization at different $t$'s for $S_z=0$ in a trapped system for $\eta=0.1$ and $\eta=0$. 
(e)  Largest level spacing $s_{\rm max}$ amongst ten lowest levels averaged over all polarization sectors, for a trapped system at $\eta=0.1$ and for $\eta=0$.}
\end{figure*}

Finally, we turn our attention to the excitation spectra. For selected $t$, 
we compare the spectra of the trapped and the homogeneous system in different polarization sectors in Figure \ref{fig:spectra}(a--d). 
For the homogeneous case, a crucial feature of these spectra has been noticed in Ref. \cite{Hauke2010}: 
Taking into account all polarization sectors, the level spacings are much more homogeneous around $t \approx 0.6$ and $t \approx 1.4$, 
than in other regimes where, in each spin sector, one (or few) strongly states are separated from higher states by a large gap. 
While in a finite system the ground state energy is different in different polarization sectors, 
one expects that in the thermodynamic limit all ground states collapse to a degenerate manifold which is separated from excited states by a gap. 
This manifold is the basis of a ``tower of states'' \cite{bernu94}, from which, by breaking of the U(1) symmetry, N{\'e}el ordered phases can arise. 
On the other hand, for the more homogeneous spectra around $t \approx 0.6$ and $t \approx 1.4$, we do not expect this mechanism to work. 
Therefore, these spectra are rather characteristic of a spin liquid phase than of an ordered phase.

To quantify this different behavior we shall look at the gap averaged over all polarization sectors. Since there might be quasi-degenerate levels, 
it is not always obvious which of the levels shall be taken as ground states, and which as excited states. For this reason, we associate the largest 
level spacing $s_{max}(S_z)$ within the ten lowest states as the energy gap. For a system with a ''tower of states'', that is with a gapped ground 
state manifold in all polarization sectors, this quantity should remain large when averaged over all polarizations. This average is shown in 
Figure \ref{fig:spectra}(e), for both a trapped and a homogeneous system. In both cases, it exhibits two clear minima near $t \approx 0.6$ and $t \approx 1.4$. 
This indicates a breakdown of order around these minima, suggesting the emergence of spin liquid behavior.

\subsection{Increasing lattice size: DMRG results \label{DMRG}}
\begin{figure*}
\includegraphics[width=1.0\columnwidth]{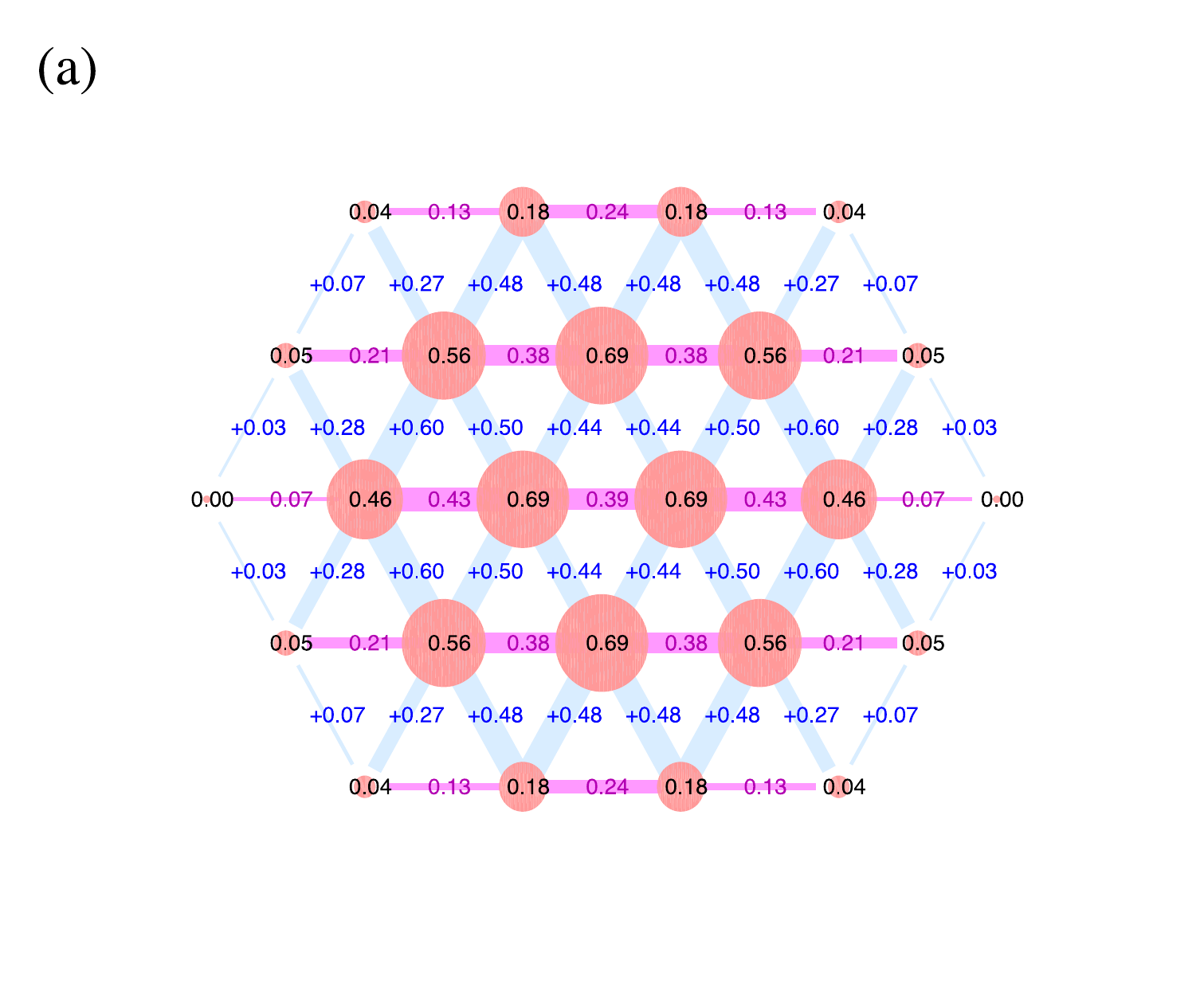} \includegraphics[width=1.0\columnwidth]{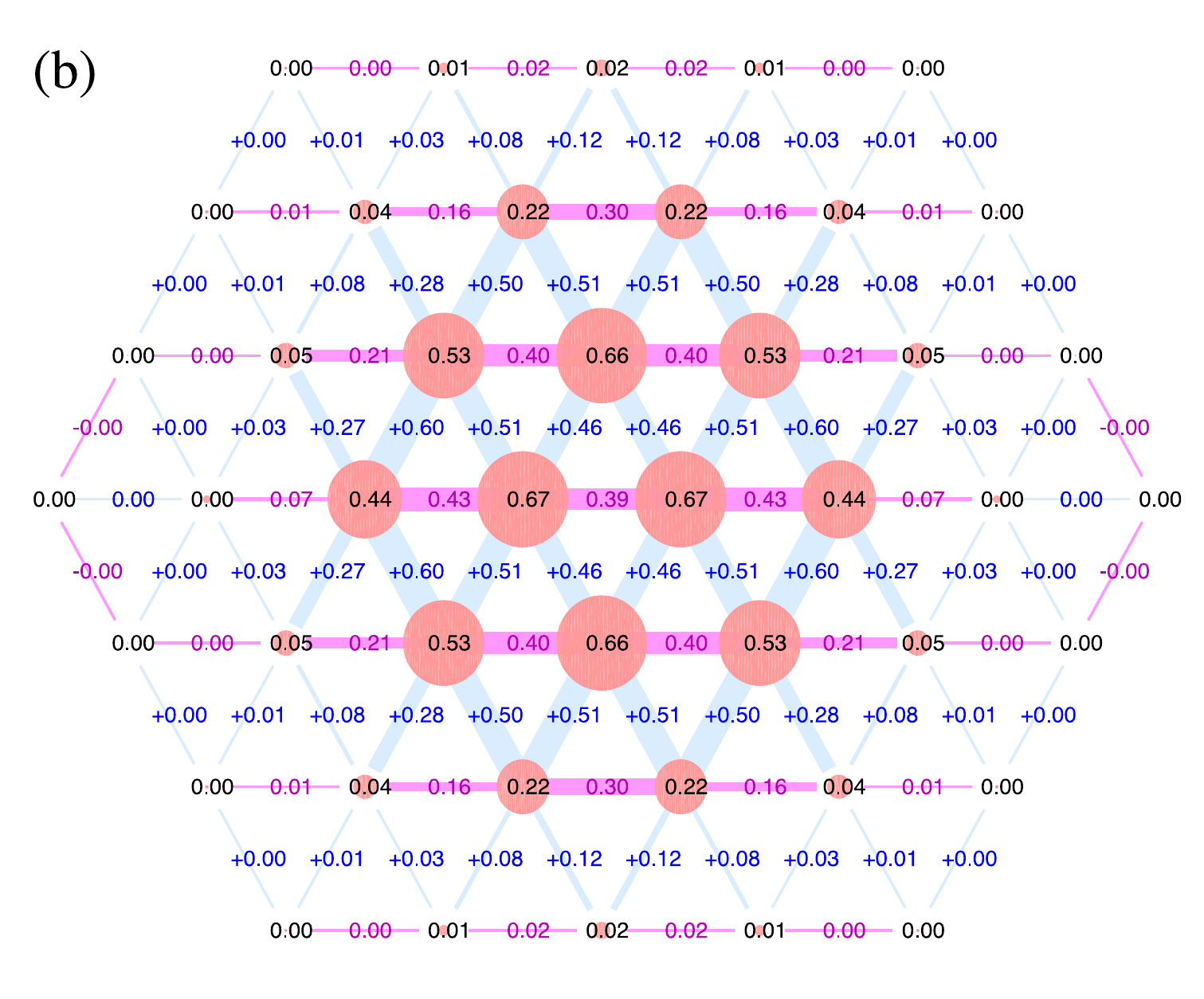}
\caption{\label{fig:dmrg}
(Color online) DMRG results. 
The density $\langle n_i \rangle$ (indicated by the area of the red circles) and coupling $\langle \hat{b}_i^{\dag} \hat{b}_j \rangle$ 
(indicated by thickness of magenta positively-valued bonds and blue negatively-valued bonds) of a trapped cloud. On the left we have a simulation of 24 sites, 
as analyzed earlier by the exact diagonalization. On the right a larger system of system of 44 sites with the same trapping 
parameters indicates that the great majority of the atoms are located in the central 24 sites, while the correlations in the centre agree to within a few percent, 
indicating that the boundary conditions have a small effect in this case. }
\end{figure*}

We finally present first results from the density-matrix renormalization group (DMRG)~\cite{white92} method, which allows us to study moderately 
larger two-dimensional systems than is accessible via exact diagonalization~\cite{stoudenmire2012}. Briefly, DMRG is a variational method that
 adaptively selects the most relevant subspace of the full Hilbert space, relative to a series of bi-partitions. This proves extremely effective 
 for large one-dimensional systems, where the so-called `area law' for entanglement entropy indicates that ground states of gapped systems have 
 bounded entanglement entropy and can be represented faithfully in DMRG calculations. In higher-dimensional systems, the method can still be 
 applied to take advantage of the area law, where in 2D systems the cost of an accurate simulation grows exponentially in the system's width, rather than its volume.

Starting with the Hamiltonian above for the trapped systems, we begin by investigate finite-size effects introduces by exact-diagonalization on small systems. 
In Fig.~\ref{fig:dmrg}, we show results for a trapped system of seven atoms with $t = 5$ and $\eta = 2.5$. 
Under these conditions, the harmonic trap confines the great majority of the atoms to the central regions studied by exact diagonalization; 
for weaker trapping parameters or greater number of atoms one would expect greater effects from the hard-wall boundary imposed by simulations of smaller systems.


 \section{Conclusions and Outlook}
 \label{Conclusion}
In this paper we have studied the fate of QSL phases in realistic experimental conditions, namely, in presence of an harmonic confinement.  
The modified spin wave theory, which was previously formulated for bosons in a triangular lattice at half filling, was re-derived for arbitrary filling factors. 
With this generalization, it can be used to capture, within a local density approximation, the physics of inhomogeneous systems. We have shown that the prediction 
of spin liquid behavior for an anisotropy $t\approx 1.65$ does not depend much on the filling factor, and should therefore survive in a trapped gas. 
This expectation was backed by results from exact diagonalization in lattices of 24 sites. 
These results support the existence of another QSL region at lower anisotropy, $t\approx 0.6$, which is not detected by MSW. 
Such discrepancy is not surprising. It is reasonable to expect that the MSW is able to detect a QSL phase only between two classical ordered phases --the QSL phase at $t\approx 1.65$
appears between spiral and 2D-N\'eel phases--  while it is blind to transitions that are purely quantum. 
One may wonder that this happens only because the optimization is done starting by the classical solution. 
In fact unbiased direct searches of global minima provided the same or more energetic metastable solutions. 
Apparently, the optimal solution of the MSW is always a deformation of the classical one: perhaps, this is not so surprising because 
the spin wave approach is an expansion in $\frac n{2S}$ and the terms in $(\frac n{2S})^2$ included in the  
MSW are corrections to the terms considered in the LSW.
The exact diagonalization approach allowed us also to go beyond the local density approximation, and to study inhomogeneities on small scales. 
On this level, we have found no essential effect due to the trap for realistic choices of the trapping frequency.
 While the finite-size corrections are certainly expected to affect the exact diagonalization results,
they should not exceed the 10-20 \%. As the observables computed are global one would argue that the QSL behavior extends at least 
to entire lattice (of 24 spins) considered.
While final-size effects are not directly visible in the MSW because we used periodic boundary conditions, 
they enter by determining the quality of local density approximation. 
Until the trap is not steep, and at the center is never so, the MSW suggests that, by taking optimal value of $t\approx 1.65$ at the center of the QSL region, 
the QSL phase should be visible even if the filling is changing considerably. Suppose, for instance, that the trap is tuned 
to have an occupation of around 3.7 atoms per site at the center, that to say 20\% above the half-filling condition. 
Then, we could conclude that if we reach an occupation of 3.3 atoms per site --20\% below half filling-- 
at 10 lattice sites or more from the center, at the same time, we are within validity of local-density approximation, in the QSL phase as predicted by MSW theory, 
and we limit the corrections due to the finite size as they are expected to go down as the inverse of the diameter
of the region considered.
Our study therefore provides strong hints for a robust spin liquid phase of bosons in anisotropic triangular lattices with antiferromagnetic tunnelings, 
which is not affected by weak trapping potentials as used in experiments.

The robustness of the spin-liquid phase in presence of a weak harmonic confinement allow for the experimental investigation of these exotic quantum phases. 
The realization of the XX Hamiltonian for bosons in the strongly correlated regime relies on the periodic driving of the triangular optical lattice, 
which allows inverting the sign of the tunneling matrix elements as well as controlling their amplitude. 
The ability to tune the tunneling amplitude independently from the on-site interaction allows reaching strongly correlated phases where $U \gg |J_{eff}|$ without 
increasing the lattice depth. Indeed, as the effective tunneling $J_{eff}$ follows a zeroth-order Bessel function as the shaking amplitude is increased, the system shall first enter a 
Mott-insulating phase before reaching the anti-ferromagnetic side of the phase diagram and thus the quantum spin liquid phase. 
Such a trajectory has allowed for a reversible crossing of the superfluid to Mott-insulator phase transition in a driven cubic lattice \cite{Zanesini09}. 
One limiting factor however are multiphoton resonances to higher lying Bloch bands, which critically reduce the coherence of the bosonic gas \cite{Weinberg15}. 
These resonances occur when a multiple of the shaking frequency matches the gap between the renormalized bands. Therefore an optimized scheme for crossing the quantum phase transition 
while avoiding such resonances has to be developed.

\begin{acknowledgments}
We acknowledge financial support from the EU grants EQuaM (FP7/2007-2013 Grant No. 323714), OSYRIS (ERC-2013-AdG Grant No. 339106), 
SIQS (FP7-ICT-2011-9 No. 600645), QUIC (H2020-FETPROACT-2014 No. 641122), Spanish MINECO grants (Severo Ochoa SEV-2015-0522 and FOQUS FIS2013-46768-P), 
Generalitat de Catalunya (2014 SGR 874), and Fundaci\'o Cellex.
\end{acknowledgments}

\end{document}